\documentclass[article]{aastex631}

\usepackage{amsmath}
\usepackage{multirow}
\usepackage{pgfplots}

\begin{document}

%\title{Exploring the nulling parameter space: new methodologies and quasiperiodic analysis}
\title{The ONs and OFFs of Pulsar Radio Emission: Characterizing the Nulling Phenomenon}

\author[0000-0001-5561-1325]{Garvit Grover}
\affiliation{International Centre for Radio Astronomy Research, Curtin University, Bentley, WA 6102, Australia}

\author[0000-0002-8383-5059]{N. D. Ramesh Bhat}
\affiliation{International Centre for Radio Astronomy Research, Curtin University, Bentley, WA 6102, Australia}

\author[0000-0001-6114-7469]{Samuel J. McSweeney}
\affiliation{International Centre for Radio Astronomy Research, Curtin University, Bentley, WA 6102, Australia}

\author[0000-0001-6840-4114]{Christopher P. Lee}
\affiliation{International Centre for Radio Astronomy Research, Curtin University, Bentley, WA 6102, Australia}

\author[0000-0001-7509-0117]{Chia Min Tan}
\affiliation{International Centre for Radio Astronomy Research, Curtin University, Bentley, WA 6102, Australia}

\author[0000-0002-9077-6026]{Shih Ching Fu}
\affiliation{International Centre for Radio Astronomy Research, Curtin University, Bentley, WA 6102, Australia}

\author[0000-0001-8845-1225]{Bradley W. Meyers}
\affiliation{International Centre for Radio Astronomy Research, Curtin University, Bentley, WA 6102, Australia}
\affiliation{Australian SKA Regional Centre (AusSRC), Curtin University, Bentley, WA 6102, Australia}




\begin{abstract}
Radio emission from pulsars is known to exhibit a diverse range of emission phenomena, among which nulling, where the emission becomes temporarily undetectable, is an intriguing one. Observations suggest nulling is prevalent in many long-period pulsars and must be understood to obtain a more comprehensive picture of pulsar emission and its evolution. One of the limitations in observational characterisation of nulling is the limited signal-to-noise, making individual pulses often not easily distinguishable from noise or any putative faint emission. Although some of the approaches in the published literature attempt to address this, they lose efficacy when individual pulses appear indistinguishable from the noise, and as a result, can lead to less accurate measurements. Here we develop a new method (the $\mathbb{N}$sum algorithm) that uses sums of pulses for better distinguishability from noise and thus measures the nulling fraction more robustly. It can be employed for measuring nulling fractions in weaker pulsars and observations with a limited number of observed pulses. We compare our algorithm with the recently developed Gaussian Mixture Modelling approach, using both simulated and real data, and find that our approach yields consistent results for generic and weaker pulsars. We also explore quasi-periodicity in nulling and measure the related parameters for five pulsars, including PSRs~J1453$-$6413, J0950$+$0755 and J0026$-$1955, for which these are also the first such measurements. We compare and contrast our analysis of quasi-periodic nulling with previously published work and explore the use of spin-down energy loss ($\dot E$) to distinguish between different types of modulation behaviour.
\end{abstract}



\section{Introduction} \label{sec:intro}

Pulsars are known to exhibit a wide-ranging emission phenomenology, many of which can be grouped under the broad category of amplitude or phase modulation phenomena. Of such, one of the most intriguing is pulse nulling, which is the phenomenon whereby the pulsar emission becomes temporarily undetectable. It was originally discovered by \cite{1970Backer}, and over the past five decades, $6\%-8\%$ pulsars have been identified to show nulling \citep[e.g.][]{Wang2007,2019Konar, Wang2020,Sheikh2021}. 
%\footnote{\url{http://www.ncra.tifr.res.in/\~sushan/null/null.html}}. 
%avoid footnote in intro; add more references e.g. large sample studies like Wang+2007, Basu+2020, etc. 

The nulling duration tends to vary greatly from pulsar to pulsar; it may range anywhere from 1--2 pulses at a time, to hundreds or thousands of pulses, and in extreme cases, may extend up to timescales of the order of weeks or even months \citep[e.g.][]{2006Kramer,2014Gajjar,2014Kerr, Basu2017}. The degree of nulling is typically quantified in terms of the so-called nulling fraction ($n_{\rm F}$), which is the fraction of pulses that null when observations are made over sufficiently long time spans. Other parameters include the nulling or burst duration, i.e. the average duration (in rotation periods) that nulling/bursting pulses tend to last, and a (quasi)periodicity in nulling (if any) -- this is the periodicity in a null-burst sequence. As with most other pulsar emission properties, nulling behaviour also
differs from pulsar to pulsar. 

Nulling, when extreme, with durations extending to hours or even days at a time,  can make it very difficult to detect such objects in large pulsar surveys, as many of them, e.g. the high-time resolution radio universe (HTRU) survey \citep{2010MNRAS.409..619K}, Green Bank North Celestial Cap (GBNCC) survey \citep{2014ApJ...791...67S} and the likes have relatively short dwell times of the order 2\,min--10\,min. Long-duration nulling may negatively impact the sensitivity of traditional periodicity searches  \citep{2024bGrover}. Such is the case with rotating radio transients (RRATs), which can be considered to be extreme cases of nulling, as they are more easily detectable via single-pulse searches \citep{2006Mclaughlin,2011Keane}.
%cite the seminal paper McLaughlin et al. 2006

Although the cause of nulling remains unknown, it seems prudent to assume it is closely linked to the physics of pulsar emission, which has been a long-standing puzzle. Understanding how pulsar emission is temporarily suppressed (or not detectable) can provide insights into the underlying high-energy physics. While some of the work suggests that nulling is due to geometric effects \citep[]{ 2005Dyks, 2007Zhang,Herfindal2009DeepBehaviour}, some others suggest that nulling originates as a result of magnetospheric phenomena occurring intrinsically in the pulsar \citep[]{1982Filippenko, 2010Timokhin,2014Gajjar}. 
%Due to the wide variety of nulling behaviours, the cause of nulling is likely an intrinsic phenomenon in the magnetosphere. 
Nulling is also observed to be generally a broadband phenomenon \citep{Basu2017, Wang2020}, although the degree of nulling may differ between different frequencies \citep[e.g.][]{2007Bhat, 2017Naidu}.

If the cessation of the pulsar emission is permanent (i.e. the pulsar is no longer detectable), it is referred to as `pulsar death'. This is thought to happen once the rotation of the pulsar becomes too slow to produce detectable radio emission \citep{Sturrock1971}, and is indicated by the so-called \emph{death lines} on the period-period derivative ($P\dot P$) diagram. 
%Due to the similarities between the temporary and permanent cessation of emission, 
Therefore, it is logical to consider nulling as an indicator of a pulsar approaching its death line. Although some of the past studies \cite[]{ 1976Ritchings, 1992Biggs, Sheikh2021}
% [cite wang, ritchings, and more form sonar, wang 2020, etc] 
have shown some positive correlations between nulling fractions and pulsar age/period, there have also been studies that have found no relationships, and some even reporting negative correlations \citep[cf.][and references therein]{1986Rankin, Wang2007, Sheikh2021}. 
Notwithstanding such a divergent picture, many of the newly found long-period pulsars (with $P \gtrsim 10$\,s) display nulling and in fact lie in proximity to the death lines \citep{Tan2018,  Morello2020discovery,Caleb2022}. On the other hand, there have also been similar objects; e.g. J1856$+$0211 \citep{Su2023} and J0311$+$1402 \citep{2025Wang}, that do not show any observational evidence for nulling. 
In short, further investigations are clearly needed in order to better understand this phenomenon and its connection to radio pulsar emission or putative death. 
%This has again brought confusion and interest to the aforementioned claims, urging for more analysis in the field. 

In this paper, we analyse a select sample of pulsars for their nulling properties, specifically the nulling fraction and quasi-periodicity. In Sections \ref{sec:classnull} to 
%, \ref{sec: comparison} and 
\ref{sec:nullfmeasurements}, we introduce a novel method for measuring the nulling fraction, specifically for weaker pulsars, and test it on our datasets. We classify and measure quasiperiodicity in nulling in Section \ref{sec:quasip_analysis} and in Section \ref{sec:discusison}, we reflect on our new method and compare its efficacy with others, and the possible ways it can be improved. We also investigate quasi-periodically in nulling and its relationship with other pulsar parameters.

\section{Motivations and Overview}
%\subsection{Nulling measurements}
\subsection{Measuring nulling parameters}

The majority of nulling studies typically report the nulling fraction, as it is often the easiest parameter to measure. So any study that involves linking nulling with other pulsar parameters relies on accurate measurements of the nulling fraction. There are two popular methods 
%in this context: 
that can be used to measure the nulling fraction: 
one of them is based on a set threshold of pulse fluence or amplitude, and the other makes use of pulse fluence distributions. The former relies on the fluence of individual pulses being well above the noise level, so that the nulling can be accurately categorised for each pulse via a singular threshold. While this appears to be analytically the most accurate method for measuring the nulling fraction, not all pulsar observations may have sufficiently high signal-to-noise (S/N) to perform such an analysis, in which case, not all pulses can be unambiguously classified into an `ON' vs `OFF' state. Consequently, its applicability is largely limited to bright pulsars or where the sensitivity is not a limitation. 

Methods that rely on pulse fluence distributions are more robust; they compare the pulse fluence distribution from the on-pulse regions to that of the off-pulse region to measure the nulling fraction. While this does not allow us to measure the nulling and/or burst durations readily, it allows for a systematic analysis of a large group of pulsars more efficiently. A popular algorithm in this category is the method originally proposed by \citet{1976Ritchings}, which relies on the expression, ON -- $n_{\rm F}\times$OFF, where ON is the distribution of the on-pulse fluence measurements, OFF is the distribution of off-pulse fluence and $n_{\rm F}$ is the nulling fraction, which is limited to $0\leq n_{\rm F}\leq 1$. It assumes that the pulse fluence distribution of nulling pulses is some scaled version of the off-pulse fluence distribution and hence returns a $n_{\rm F}$ that minimises the above expression for pulse fluences $\leq0$. I.e. it assumes that all contributions to the fluence below zero are from nulling.

A more recent method is the Gaussian mixture model (GMM) method developed by \citet{Kaplan2018},  which aims to improve upon Ritchings' method and yield more robust measurements. GMM decomposes the on-pulse histogram of pulse fluence into X+1 Gaussians, where +1 refers to the Gaussian representing the noise-like fluence of nulling pulses. The contribution of the noise-like Gaussian of nulling pulses to the on-pulse histogram is the $n_{\rm F}$. \cite{Kaplan2018} showed that Ritchings' method overestimates the nulling fraction for weaker pulsars, and hence, GMM is likely to be more accurate for such pulsars. GMM was further improved by \cite{Anumarlapudi2023}, with additional capability of fitting Gaussian distributions with exponential tails for skewed pulse fluence distributions. It has been used in a number of recent studies \citep{Grover2024a,Tedlia2024, Xu2024, 2025McSweeney}. However, even this approach seems to give less accurate results for pulsars where single pulses are not easily distinguishable from the noise \citep[e.g.][]{2025McSweeney}. In essence, a detailed scrutiny of nulling characterisation, particularly for extracting reliable measurements for weaker pulsars, and/or when observations are limited by signal-to-noise per pulse, 
is necessary in order to make further progress.  
%Hence, alternative solutions should be explored to measure nulling in very weak pulsars.
By using multiple robust methodologies, one can obtain accurate nulling fraction measurements for pulsars of a larger variety of brightnesses, allowing us to be less biased in the selection of pulsars. Remeasuring $n_{\rm F}$ with new methodologies will allow us to test correlations between the parameters and pulsar rotation period/age more accurately.

%\subsection{Periodic nature of nulling}
\subsection{Quasi-periodicity in nulling}

Despite \cite{1970Backer}'s original discovery reporting that the occurrence of nulls displays some periodicity, the general perception has been that the occurrence of nulls is a stochastic process that is not necessarily periodic and varies between pulsars.  Some later studies \citep{Wang2007,2007Herfindal,2008Rankin,Herfindal2009DeepBehaviour} provided further observational evidence in support of periodicity in nulling, and more recently, there have been a number of investigations that further explored this aspect of nulling \citep[e.g.][]{Basu2017,Basu2020,Wang2020, Anumarlapudi2023,Grover2024a, Tedlia2024, Xu2024}. 

The cause of this quasi-periodicity remains unclear; however, nulling is also often observed in conjunction with sub-pulse drifting in pulsars \citep{Weltevrede2006,2007Herfindal,Herfindal2009DeepBehaviour}, thereby suggesting that these two phenomena are possibly linked. Sub-pulse drifting is generally explained in terms of a rotating carousel of beamlets near the magnetic pole of the pulsar \citep{Ruderman1974}.
%\cite{Herfindal2009DeepBehaviour,2007Herfindal} 
\citet{2007Herfindal} suggested quasiperiodicity may arise due to `pseudo-nulls', i.e. nulls occurring due to gaps between beamlets in the carousel. As the carousel of beamlets rotates periodically around the magnetic pole, this theory may naturally explain the observed periodicity in nulling. In a more recent work, \cite{Grover2024a} compiled a list of 33 quasi-periodically nulling pulsars, some of which also showed sub-pulse drifting, and demonstrated that there was no correlation between quasi-periodicity in nulling and sub-pulse drifting parameters. Furthermore, not all pulsars that showed quasi-periodic nulling show clear signs of sub-pulse drifting, suggesting there may not be a connection between the two phenomena.

There have also been studies to explore the possible links between quasiperiodic nulling and other pulsar parameters. For instance, \citet{Wang2007} found no significant correlation between the nulling periodicity and pulsar period or characteristic age; however, their data were time-averaged (by up to 10\,s) in order to improve S/N, which can lead to inaccurate measurements. More recent work by \citet{Basu2017,Basu2020} suggests that quasiperiodically nulling pulsars are clustered in a distinct region when the nulling period is plotted against the spin-down energy loss ($\dot E$), and thus in the diagram, appear separated from pulsars that show other types of periodicities (e.g. sub-pulse drifting). Although the pseudo-nulling explanation from \cite{2007Herfindal} and \cite{Herfindal2009DeepBehaviour} implies nulling to be a geometric effect caused by sub-pulse drifting, \citep{Basu2020}'s work shows a distinct separation in the population of sub-pulse drifting pulsars and quasiperiodically nulling pulsars. Additionally, a relationship with $\dot E$ would imply quasiperiodic nulling to be more than a geometric effect but one related closely with the magnetic/global evolution of the pulsar.

\cite{Grover2024a} collated data from \cite{Basu2020} and \cite{Anumarlapudi2023}, and they find that the vast majority of pulsars that exhibit quasiperiodic nulling lie in the death valley in the $P\dot P$ diagram. This trend was further strengthened by more recent works by \cite{Xu2024} and \cite{Tedlia2024}. It appears that these recent studies are suggestive of an age or $\dot E$ threshold for pulsars to show a quasi-periodicity in nulling, implying an evolution of nulling with time. Interestingly, this population lies in the region where pulsars are theorised to stop emitting radio radiation, and hence it is possible that the nulling quasiperiodicity may be related to the evolution of pulsar emission.

To explore these hypotheses further, larger population studies should be conducted by large-scale surveys to systematically measure nulling and hence nulling quasiperiodicity to provide a larger sample size. In this paper, we introduce a methodology for measuring and categorising quasiperiodicity, apply it to the pulsars in our sample, and expand the number of quasiperiodically nulling pulsars.

\begin{deluxetable*}{lccccc}
\tablecaption{A list of the pulsars used in this study, the SMART survey observations from which they were taken, and other such properties. }\label{tab:pulsars}
\tablehead{
\colhead{Obs} & \colhead{PSRJ} & \colhead{$P$} & \colhead{$\dot P$} & \colhead{DM} & \colhead{MJD} \\
 &  & \colhead{(s)} & ($10^{-15}$) & \colhead{(cm$^{-3}$ pc)} & 
}
\startdata
B07 & J0026$-$1955 & 1.31 &$0.44$ & 20.87 & 58434\\
R04  & J0452$-$3418  & 1.66  & $2.78$ & 19.78  & 58757 \\
G02  & J0630$-$2834  & 1.24  & $7.19$ & 34.63  & 58841 \\
G02  & J0742$-$2822  & 0.16 & $16.8$ & 73.75  & 58841 \\
G06  & J0837$-$4135  & 0.75 & $3.54$ & 147.21 & 58900 \\
G11  & J0953$+$0755  & 0.25 & $0.23$ & 2.94  & 58909 \\
P02  & J1136$+$1551  & 1.19  & $3.73$ & 4.78  & 59301 \\
P02   & J1239$+$2453  & 1.38  & $0.96$ & 9.25  & 59301 \\
P07  & J1430$-$6623  & 0.79 & $2.78$ & 65.12 & 59314 \\
P07   & J1453$-$6413  & 0.18  & $2.74$ & 71.18  & 59314 \\
P13  & J1456$-$6843  & 0.26 & $0.10$ & 8.63  & 60069 \\
P14  & J1559$-$4438  & 0.26 & $1.02$ & 56.09  & 60070 \\
O01  & J1645$-$0317  & 0.39 & $1.78$ & 35.76  & 60076 \\
O06  & J1901$-$0906  & 1.78  & $1.64$ & 72.68  & 60094 \\
O06 & J1913$-$0440  & 0.83  & $4.07$ & 89.39  & 60094 \\
O06 & J1919$+$0021  & 1.27  & $7.68$ & 90.32  & 60094 \\
O06 & J1945$-$0040  & 1.05  & $0.53$ & 59.27  & 60094 \\
O13 & J2046$-$0421  & 1.55 & $1.47$ & 35.94  & 60118 \\
\enddata
% \tablecomments{Column descriptions: }
\end{deluxetable*}

%\section{Observations and data}
% why not just Data given no new observations are involved? 
\section{Data}
\label{sec:obs}
% \begin{itemize}
%     \item SMART survey
%     \item MWA
%     \item low-frequency drift scan 
% \end{itemize}

The data used for this project were collected as part of the Southern-sky MWA Rapid Two-meter (SMART) survey, which uses the Murchison Widefield Array \citep[MWA;][]{2013Tingay,2018Wayth}, a low-frequency interferometer telescope located at Inyarrimanha Ilgari Bundara, the Commonwealth Scientific and Industrial Research Organisation (CSIRO) Murchison Radio-astronomy Observatory. For details on the pulsars and observations used in this study, see Table \ref{tab:pulsars}.
SMART observations are 80-minute long drift scans, and the survey covers the whole southern sky at declinations $<30^\circ$ in the frequency band 140--170\,MHz. For more details on the SMART survey data collection, see \cite{2023Bhat_a, 2023Bhat_b}. As the MWA voltage capture system (VCS) underwent an upgrade during the course of the SMART survey, the data used for this work were recorded using two different systems, i.e. the legacy VCS \citep{2015Tremblay} and MWAX VCS \citep{2023Morisson}. Further details about the calibration and processing of pulsar detections in SMART will be detailed in future publications (Lee~et~al. submitted; Bhat~et~al. in~prep.).

The SMART dataset allows us to explore the nulling phenomenon at rather uncommonly observed frequencies for single-pulse work. Furthermore, much longer durations of the SMART observations (compared to other surveys) allow us to meaningfully characterise the nulling parameters at these frequencies.
However, the MWA is fairly modest in sensitivity for single-pulse studies, with a mean S/N per pulse $<<$ 1 for most pulsars. But this may be advantageous for our work as it effectively emulates the low S/N regime that we would like to test our new approach and make reliable measurements of nulling fraction and other parameters.

The pulsars chosen for this analysis were all reported to show some level of nulling. We have attempted to choose pulsars such that they have a variety of nulling fractions, pulse shapes DMs, and S/Ns. As we are limited by the coverage of the available SMART survey data, we have only processed pulsars primarily in the southern hemisphere. 
% Unintentionally, the pulsars chosen lie primarily near the central cluster of the $P\dot P$ diagram. 

\section{Analysis}
\label{sec:analysis}

\subsection{Characterising Nulling: $\mathbb N$sum Method}
\label{sec:classnull}

To tackle the challenge of measuring nulling fractions for low S/N pulsar observations, we developed the $\mathbb N$sum approach, which is a generalisation of the Gaussian mixture model \citep{Kaplan2018}. It is often the case for weak pulsars that the individual pulses show no discernible emission, and hence they can be confused for nulling pulses. However, when multiple of said pulses are summed, they do in fact show emission. We aimed to use this summation to make emissions more distinguishable from noise and, hence, measure nulling fractions more accurately.

The algorithm entails creating the fluence distribution of $\mathbb Y$ random sums of $\mathbb N$ pulses within the on-pulse region ($\mathcal{F}^{\mathbb  N}_{\mathbb Y}$); this can include bursting and nulling pulses. This distribution of fluences from random sums is a mixture of the distributions of fluences from the possible combinations of sums of pulses. The probability of picking a random nulling pulse is $n_{\rm F}$ and picking a random bursting pulse is $1-n_{\rm F}$. Hence, the probability of any combination of pulses (e.g. two bursting pulses or a nulling and a bursting pulse), no matter the $\mathbb N$, is dependent on $n_{\rm F}$; by using statistical sampling techniques (e.g. Markov-Chain Monte-Carlo; MCMC), the nulling fraction can be estimated. 

% Following an example probability tree for $\mathbb N =2$, as shown in Figure \ref{fig:probabilitytree}, the probability for each outcome can be known, and hence, their contribution (weights) to the final distribution of sums. This example demonstrates sampling with replacement and can easily be propagated for higher $\mathbb N$. The probability for each outcome is dependent on the nulling fraction $n_{\rm F}$; by using statistical sampling techniques (e.g. Markov-Chain Monte-Carlo; MCMC), the nulling fraction can be measured. 

\begin{figure}
    \centering
    \includegraphics[width=0.5\linewidth]{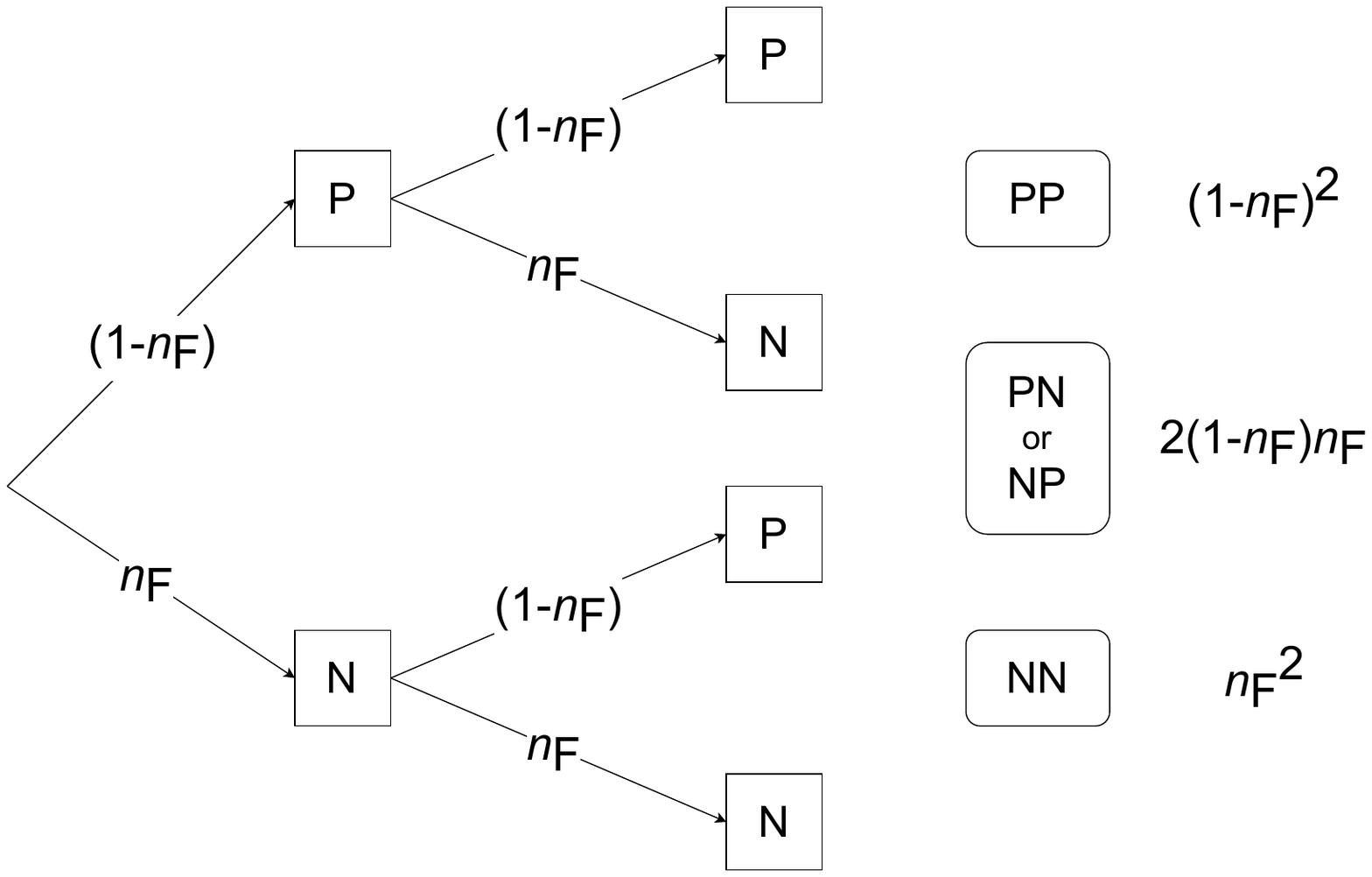}
    \caption{A probability tree of picking two random pulses ($\mathbb N = 2$), assuming replacement of samples. The right-hand side shows the possible outcomes and their probabilities.}
    \label{fig:probabilitytree}
\end{figure}

For a detailed explanation, consider the following example where the fluence of two random pulses are summed, ${\mathbb  N}=2$, and we generate a distribution with 1000 random sums ($\mathbb Y=1000$), $\mathcal{F}^{\mathbb N=2}_{\mathbb Y=1000}$. As shown in Figure \ref{fig:probabilitytree}, the probability of randomly summing two bursting pulses $\mathcal{P}_{\rm PP}=(1-n_{\rm F})^2$, two nulling pulses $\mathcal P_{\rm NN}=n_{\rm F}^2$, and one bursting and one nulling pulse $\mathcal P_{\rm PN}=2(1-n_{\rm F})n_{\rm F}$. For a general case, the probability of each scenario can be determined by:
\begin{equation}
    \mathcal P(z;{\mathbb N}) = \binom{\mathbb N}{z} n_{\rm F}^{z}(1-n_{\rm F})^{({\mathbb N}-z)}
\end{equation}\label{eq:comb}
where $z$ is the number of nulling pulses in the random sum up to a maximum value of $\mathbb N$.

Given that ${\mathbb  N}=2$ and the probabilities of the outcomes are as listed above, we can define the distribution of the fluences for each scenario of sums. For this approach, we assume the pulse fluences and the noise are drawn from a Gaussian distribution, which we call $\mathcal{N}(\mu_{\rm P}, \sigma_{\rm P})$ and $\mathcal{N}(\mu_{\rm N}, \sigma_{\rm N})$, respectively. The distribution of the sum of two independent random variables is the convolution of their respective distributions. The convolution of any two Gaussian-distributed random variables is another Gaussian random variable; its parameters are
\begin{equation}\label{eq: convolution}
    \mathcal{N}(\mu_{1}, \sigma_{1}) \ast \mathcal{N}(\mu_{2}, \sigma_{2}) = \mathcal{N}(\mu_{1}+\mu_{2}, \sqrt{\sigma_{1}^2+\sigma_{2}^2}).
\end{equation}
Hence, we can obtain distributions:
\begin{equation}\label{eq:distributions}
    \mathcal{N}(2\mu_{\rm P}, \sqrt{2}\sigma_{\rm P}),\\
    \quad\mathcal{N}(2\mu_{\rm N}, \sqrt{2}\sigma_{\rm N}),\\
    \quad\mathcal{N}(\mu_{\rm P}+\mu_{\rm N}, \sqrt{\sigma_{\rm P}^2+\sigma_{\rm N}^2}).
\end{equation}

\begin{figure}
    \centering
    \begin{minipage}{0.4\textwidth}
        \centering
        \includegraphics[width=\textwidth]{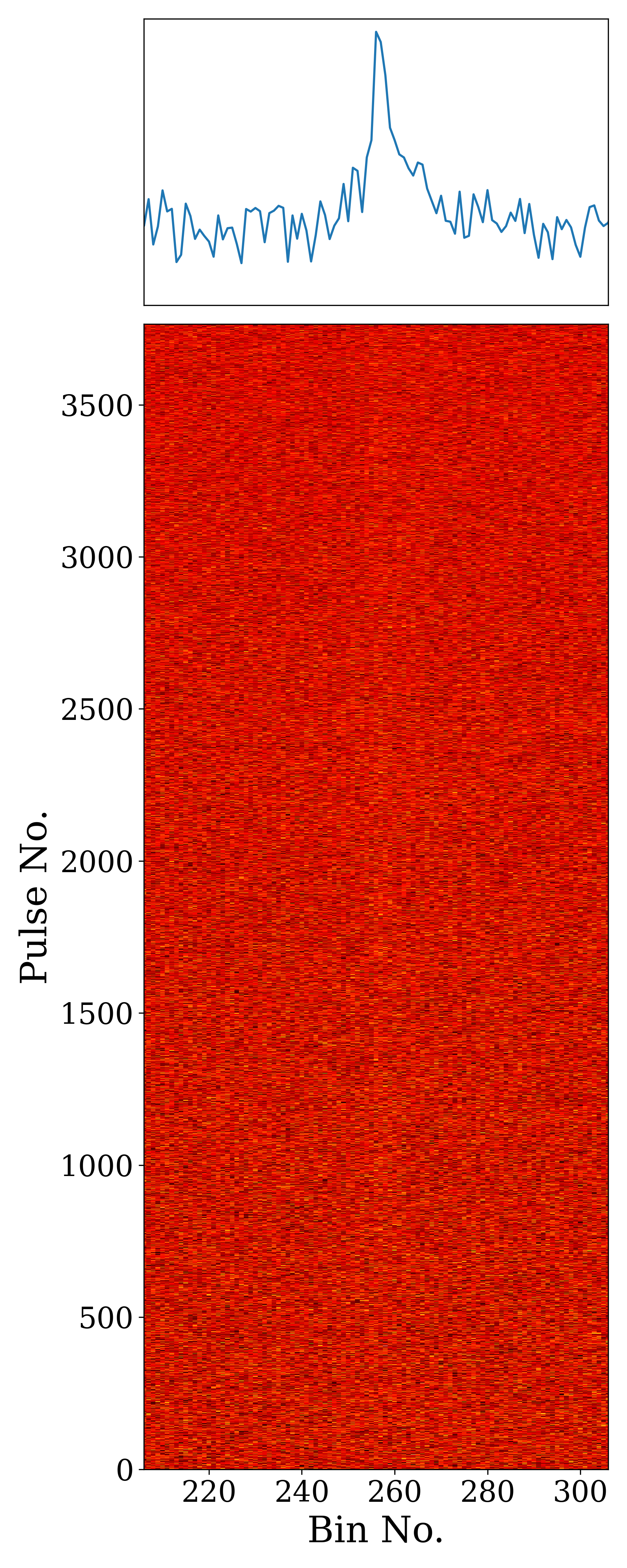}
        
    \end{minipage}%
    \begin{minipage}{0.55\textwidth}
        \centering
        \includegraphics[width=\textwidth]{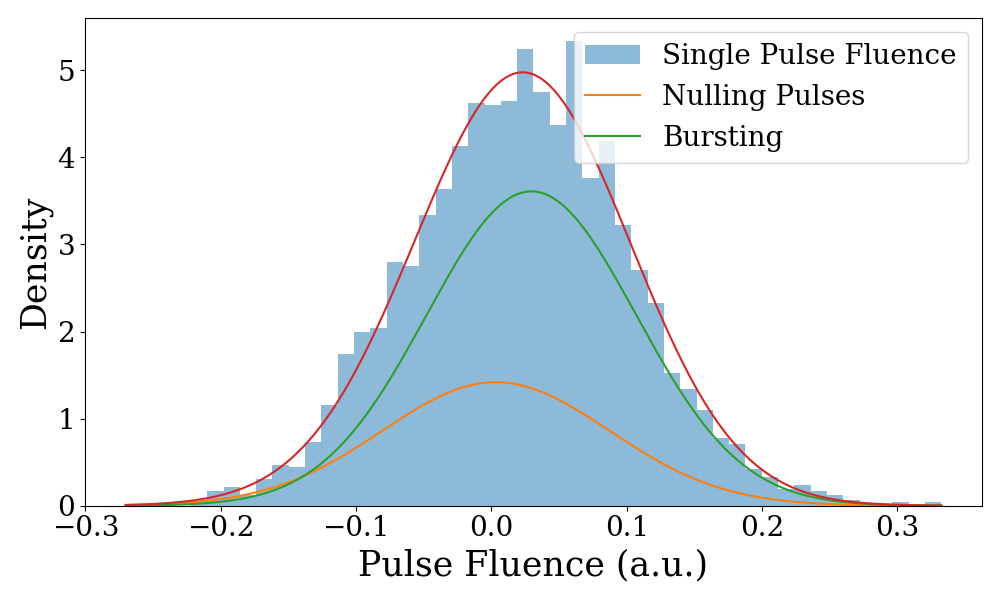}

        \vspace{0.5cm} % Space between the two right column images
        
        \includegraphics[width=\textwidth]{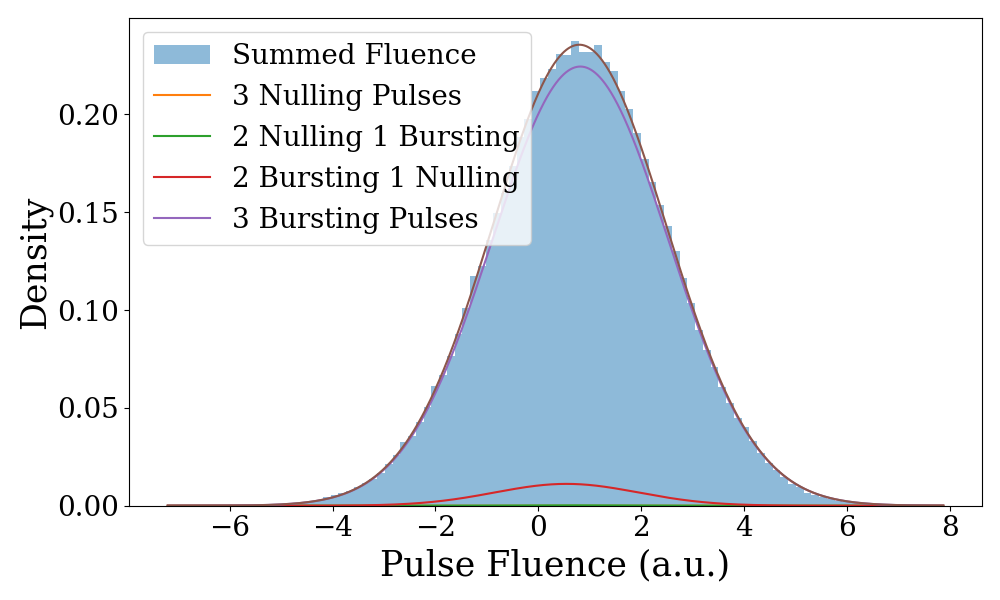}
        \vspace{0.5cm}
        \includegraphics[width=\textwidth]{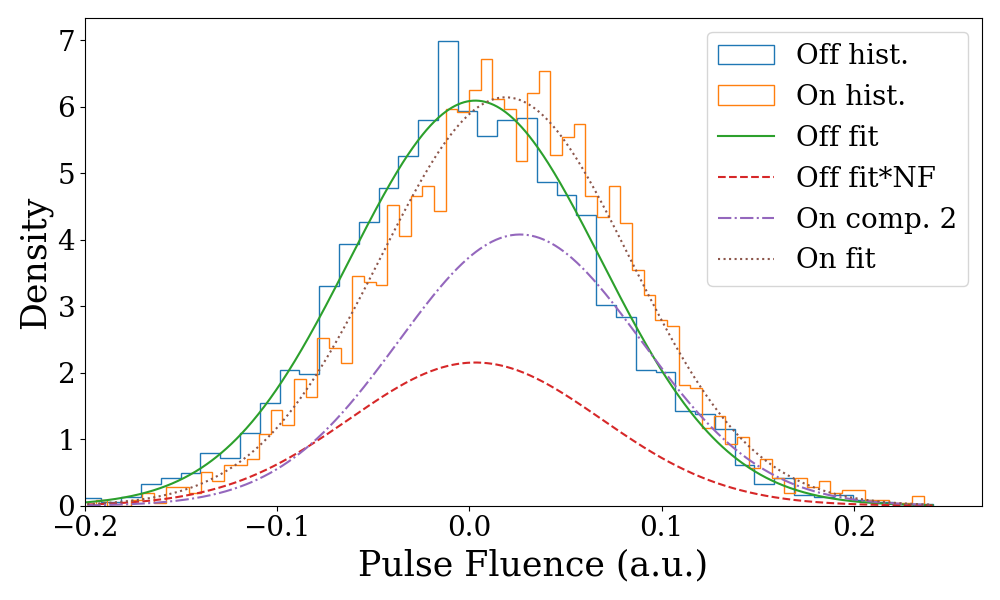}
    \end{minipage}
    \caption{On the left is the pulse stack for J1919$+$0021 in the MWA SMART observations. The top right panel shows the pulse fluence for the on-pulse regions fit with a mixture of two Gaussians, for the bursting and nulling pulses. This plot represents the $\mathbb N=1$ case where no pulses are summed. The middle right plot shows the histogram of the fluences of the sums of three random pulses ($\mathbb N=3$). As summing three pulses can have four possible outcomes, the histogram is fit with a mixture of four Gaussians. The bottom right plot is a similar histogram to those above and a direct output of \texttt{pulsar\_nulling}\footnote{https://github.com/AkashA98/pulsar\_nulling}\citep{Anumarlapudi2023}.}
    \label{fig: J1919+0021_example}
\end{figure}

\begin{figure}
    \centering
    \begin{minipage}{0.4\textwidth}
        \centering
        \includegraphics[width=\textwidth]{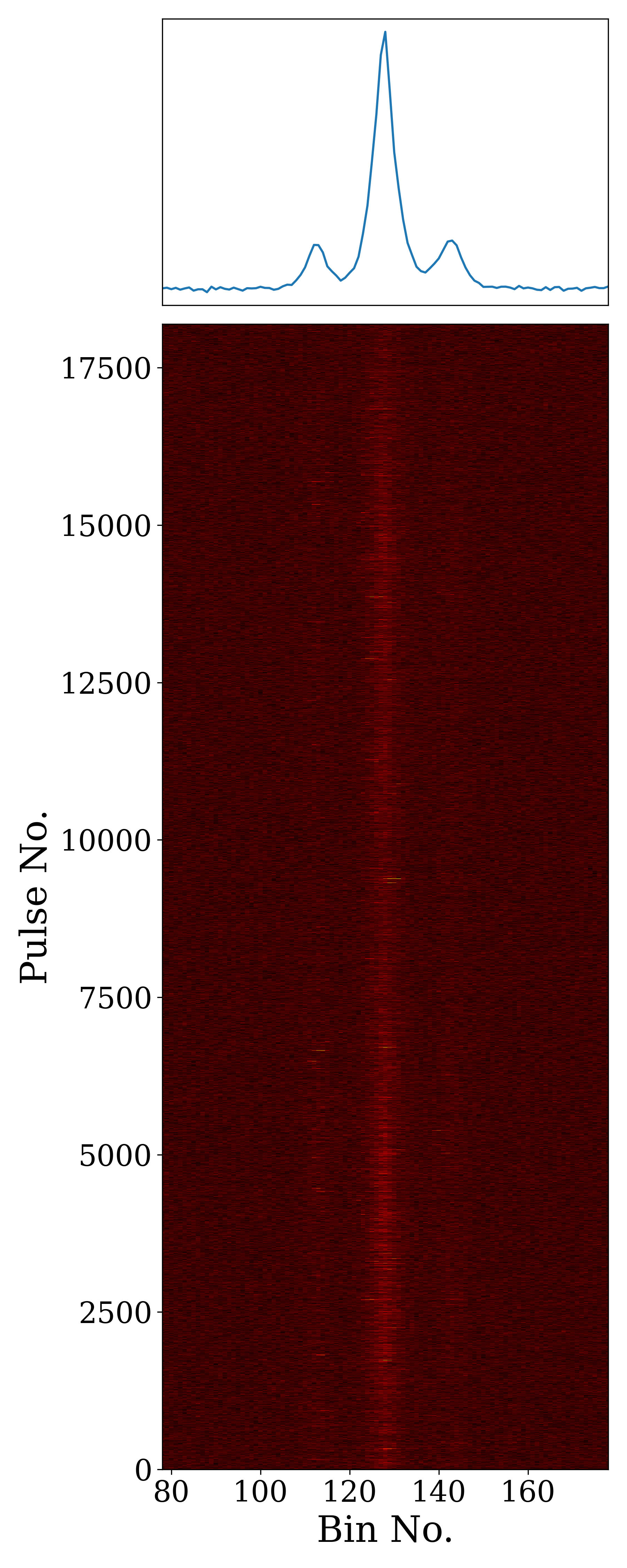}
        
    \end{minipage}%
    \begin{minipage}{0.55\textwidth}
        \centering
        \includegraphics[width=\textwidth]{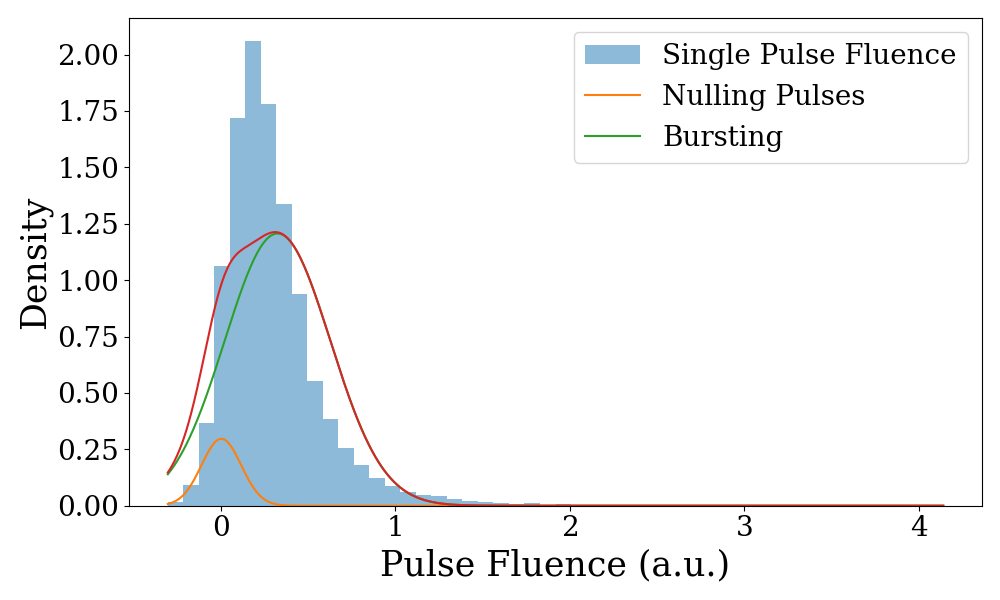}

        \vspace{0.5cm} % Space between the two right column images
        
        \includegraphics[width=\textwidth]{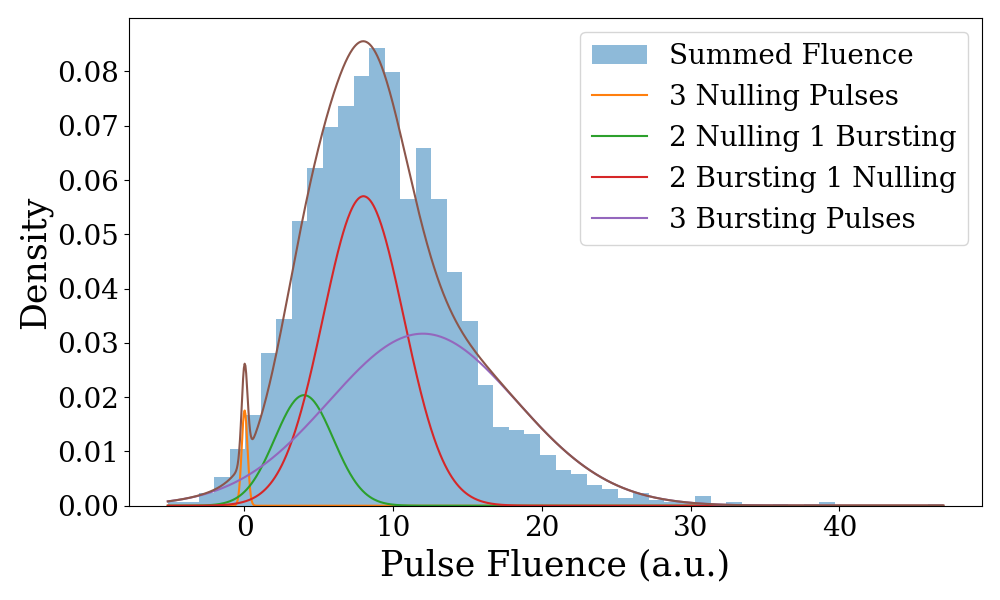}
        \vspace{0.5cm}
        \includegraphics[width=\textwidth]{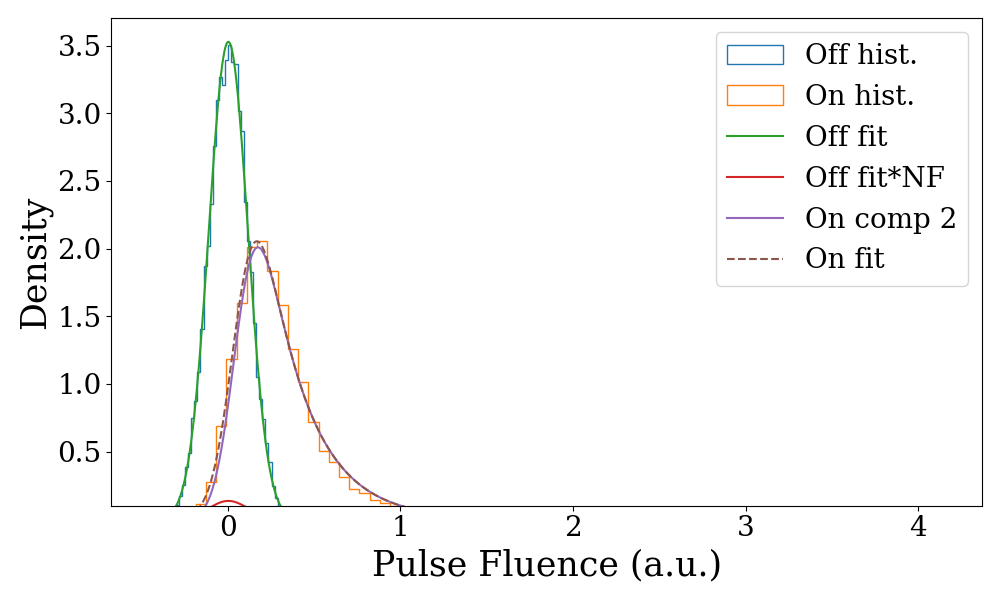}
    \end{minipage}
    \caption{Same as Figure \ref{fig: J1919+0021_example}, but for J1456$-$6843.}
    \label{fig: J1456-6843_example}
\end{figure}

Once we describe the distribution of the sum of any two pulses, then we can fully decompose $\mathcal{F}^{\mathbb N=2}_{\mathbb Y=1000}$ into a mixture of three normal distributions, with the weights of the distributions described by Equation 1 and, more directly, the right-hand side of Figure \ref{fig:probabilitytree}. 
% depending on $n_{\rm F}$

% We can assume $\mathcal{N}(\mu_{\rm N},\sigma_{\rm N})$ is the same as the off-pulse noise; however, we do not know $\mathcal{N}(\mu_{\rm P},\sigma_{\rm P})$. 
% Therefore, using MCMC, we can estimate $\mu_{\rm P}, \sigma_{\rm P}$ and $n_{\rm F}$. 
Following Bayesian inference, the likelihood to be minimised is the difference between $\mathcal{F}^{\mathbb N=2}_{\mathbb Y=1000}$ and the Gaussian mixture of the expressions in Equation \eqref{eq:distributions}. The distributions themselves are described by $\mu_{\rm N},\sigma_{\rm N}, \mu_{\rm P}$ and $\sigma_{\rm P}$. The weights of these distributions are dependent on $n_{\rm F}$. It is reasonable to assume $\mu_{\rm N}$ and $\sigma_{\rm N}$ are very similar to that of the off-pulse noise, as was done by \citep{Kaplan2018}; however, GMM considers the off-pulse noise as initial guesses for $\mu_{\rm N}$ and $\sigma_{\rm N}$ and uses MCMC to fit for all five parameters. At very small $\mu_{\rm P}$,  $\mu_{\rm N}$ and $\sigma_{\rm N}$ can be overestimated when using $\mathcal{F}$ for the likelihood, which leads to an overestimation of $n_{\rm F}$. Hence, for $\mathbb N$sum, we only consider $\mu_{\rm P},\sigma_{\rm P}$ and $n_{\rm F}$ to be free parameters . Given appropriate priors, e.g. $\mu_{\rm P}>0,\sigma_{\rm P}>0$ and $0<n_{\rm F}<1$, we use MCMC to estimate these parameters.

$\mathbb N$ can be adjusted; a higher $\mathbb N$ leads to the stretching of $\mathcal{F}^{\mathbb N}_{\mathbb Y}$ in the positive direction in the presence of emission. E.g. consider $\mu_{\rm P}>0, \mu_{\rm N}\sim 0$.  If no bursting pulses are present ($n_{\rm F}=100\%$), the mean of $\mathcal F_{\mathbb Y}^{\mathbb N}$ will remain $\sim0$ , with the variance increasing by $\sqrt{\mathbb N}$. Alternatively, if $n_{\rm F}=0\%$, the mean of $\mathcal F_{\mathbb Y}^{\mathbb N}$ increases by $\mathbb N \times \mu_{\rm P}$, making the mean more positive with larger $\mathbb N$. Therefore, in the case of weak pulsars, a larger $\mathbb N$ should provide a stronger indication of bursting pulses, if they exist, via a positive mean. Moreover, as there are more components dependent on $n_{\rm F}$ with increasing $\mathbb N$, we get a more accurate measurement of $n_{\rm F}$. 

An example of one of our fits is present in Figure \ref{fig: J1919+0021_example}, where we show the pulse fluence histograms for the $\mathbb N=1$ case, $\mathbb N=3$ case and GMM for pulsar J1919$+$0021. As is obvious from the left-hand side plots, the pulsar observation is very weak with an S/N $\sim 13$. The top right ($\mathbb N=1$) and bottom right (GMM) indicate a significant nulling fraction; while the middle right plot shows the $\mathbb N=3$ product, which indicates a smaller nulling fraction, see Table \ref{tab: nftable} for exact values. The positively shifted mean in the $\mathbb N=3$ case makes the abundance of emission in the observation readily apparent. 
% top right plot of the Figure, the pulse fluence appears to be very weak and log-normal-like for ${\mathbb N}=1$; for ${\mathbb N}=3$, the bottom left plot, the fluence histogram is more Gaussian, and shifted to the right, making a clear distinction of the abundance of emission.

As we rely on random sums, the larger the $\mathbb Y$ means we can accurately sample $\mathcal{F}^{\mathbb N}_{\mathbb Y}$, hence, allowing us to estimate $n_{\rm F}$ accurately. However, $\mathbb Y$ is limited by the number of pulses observed ($\mathbb O$) and $\mathbb N$, such that the maximum number of available sums is $\binom{\mathbb O}{\mathbb N}$. For example, for a long-period pulsar with $\mathbb O=100$ and ${\mathbb N}=2$, we can obtain ${\mathbb Y}=4950$ random sums. Hence, even with a limited number of pulses, we can obtain a relatively sufficient sample size to create a detailed $\mathcal{F}^{\mathbb N}_{\mathbb Y}$.

The downside of having a large $\mathbb N$ or performing a larger number of sums ($\mathbb Y$) is the computational costs. 
% Very large $\mathbb N$ may also lead to degeneracies as the $\mathcal{F}^{\mathbb N}_{\mathbb Y}$ may become too wide to precisely or accurately measure $n_{\rm F}$. 

Another variable is the distribution of the intrinsic fluences of the emission pulses. Typically, pulse fluences are observed to follow a log-normal distribution \citep{2012Burke-Spolaor}. Low S/N pulsars still appear to have normally distributed pulse fluences. Here, we have assumed all pulse fluences follow a Gaussian distribution as it is more analytically tractable. Additionally, we assume that pulse fluences are only drawn from one distribution; however, this may not be applicable to all pulsars. In Section \ref{sec:discusison}, we discuss our efforts in fitting log-normal distributions, which may yield more appropriate results for certain pulsars. As we solve for $\mu_{\rm P}$ and $\sigma_{\rm P}$, the ${\mathbb N}$sum approach also allows for the interpretation of intrinsic pulse fluences and how they may change between nulling pulsars.

\subsection{Comparison with Simulated Data}\label{sec: comparison}

We tested our method on simulated data to test the accuracy and reliability of the $\mathbb N$sum approach and compare its performance with the Gaussian mixture  as implemented by \cite{Anumarlapudi2023}. We simulated MWA-like noise, $\mathcal{N}(\mu_{\rm N}=0, \sigma_{\rm N}=0.3)$, and assumed intrinsic pulse energies were drawn from a normal distribution of $\sigma_{\rm P} = 0.1$, with varying $\mu_{\rm P}$ ($0.001 - 1$), similar to \cite{Kaplan2018}. We randomly sampled from the noise and intrinsic energy distributions and combined them to mimic pulse stacks of 1000 pulses with nulling fractions of 10\%, 30\%, 50\%, 70\% and 90\%. All MCMC processes had 200 initial steps and 5000 steps. For $\mathbb N = 3$, we used $\mathbb Y=5\times10^4$ sums.

\begin{figure}
    \centering
    \includegraphics[width=0.9\linewidth]{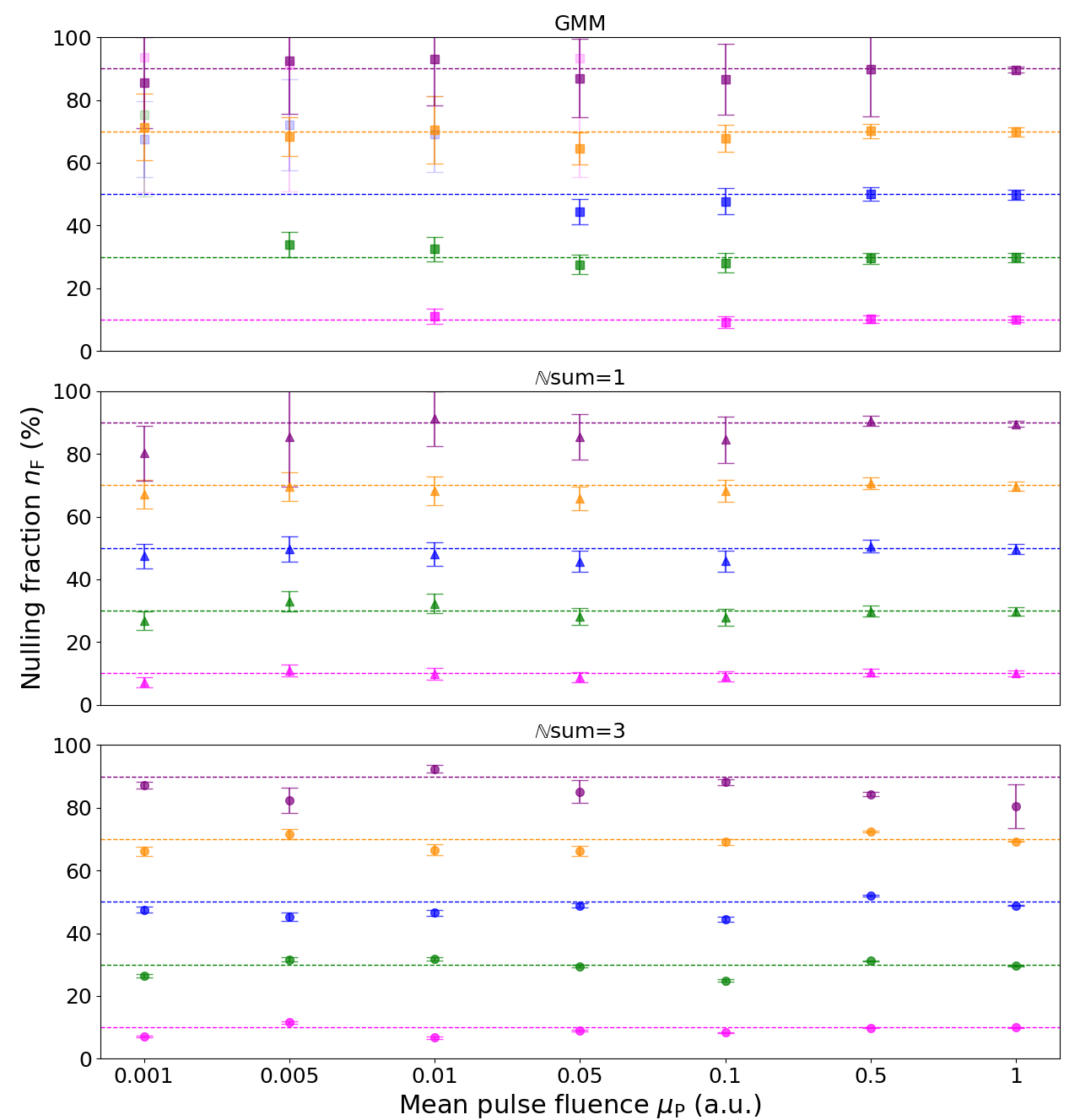}
    \caption{Plots showing the measured nulling fraction values from GMM (top), $\mathbb N=1$ (middle) and $\mathbb N=3$ (bottom) cases for simulated pulses with increasing mean pulse fluence. The noise in the simulated data was kept constant (see details in the text). The coloured dashed lines represent the injected $n_{\rm F}$ and the data points represent the measurements of the data with the injected $n_{\rm F}$. }
    \label{fig:nf_comp}
\end{figure}

Using our simulated data, we performed the $\mathbb N$sum approach with ${\mathbb N}=1$ and ${\mathbb N}=3$. Table \ref{tab: simset} and  Figure \ref{fig:nf_comp} show the expected and measured nulling fractions using the simulated data for increasing mean pulse fluence. The simulated tests show that a majority of the $\mathbb N=3$ measurements jitter around the expected value; this effect is present throughout all $\mu_{\rm P}$. In 94\% of $\mathbb N = 3$ cases, the residual error is $<$ 5\%, consistent with the expected values; however, only 20\% of the estimated values agree with the expected value within error.

% In 3 other cases, the (large) error was due to the fact that the posteriors of the MCMC analysis were bimodal, with only one ``island'' of solutions being centred on the correct values, and with the fitted values being erroneously reported as the mean of the total posterior distribution. In such cases, simply rerunning the MCMC analysis was sufficient for it to converge on the correct solution---it never converged on the incorrect ``island''.

% Although the underestimation/overestimation is not a large margin for most measurements, in some cases, the differences can be a large margin. However, even with this jitter, $\mathbb N=3$ measures the expected $n_{\rm F}$ within 5\% majority of the time.

There were three cases where $\mathbb N=3$ produced bimodal posteriors; with only one ``island'' of solutions being centred on the correct values, and with the fitted values being erroneously reported as the mean of the total posterior distribution. In such cases, simply rerunning the MCMC analysis was sufficient for it to converge on/near the correct solution (see an example of the correctly converged solution in Figure \ref{fig:nsumcorner} in Appendix) ---it never converged on the incorrect ``island''.  Similarly, there were two cases where it converged on the incorrect solution, but again, reprocessing the MCMC analysis led to expected results.

\begin{deluxetable}{ccccc}
\label{tab: simset}
\tablecaption{The comparison between the expected nulling fractions and the measured nulling fractions from GMM and $\mathbb N$sum ($\mathbb N=1$ and $ \mathbb N=3$) for a simulated dataset.}
\tablehead{
\colhead{Mean} & \colhead{Expected} & \colhead{GMM}& \colhead{$\mathbb N=1$}& \colhead{$\mathbb N=3$}\\
 \colhead{Fluence}& \colhead{$n_{\rm F} \,(\%)$}& \colhead{$n_{\rm F}\,(\%)$}& \colhead{$n_{\rm F} \,(\%)$}& \colhead{$n_{\rm F} \,(\%)$}
}
\startdata
0.001 & 10 & $93.59 \pm 42.91^b$ & $7.22 \pm 1.69$ & $7.12 \pm 0.26$ \\
0.001 & 30 & $75.33 \pm 26.20^b$ & $26.88 \pm 2.97$ & $26.37 \pm 0.55$ \\
0.001 & 50 & $67.61 \pm 12.07^b$ & $47.39 \pm 3.82$ & $47.54 \pm 0.89$ \\
0.001 & 70 & $71.38 \pm 10.66$ & $67.20 \pm 4.51$ & $66.21 \pm 1.47$ \\
0.001 & 90 & $85.51 \pm 14.56$ & $80.21 \pm 8.67$ & $87.21 \pm 1.11$ \\
0.005 & 10 & $91.14 \pm 40.38^b$ & $10.89 \pm 1.93$ & $11.53 \pm 0.29$ \\
0.005 & 30 & $33.82 \pm 4.03$ & $32.99 \pm 3.14$ & $31.64 \pm 0.64$ \\
0.005 & 50 & $72.15 \pm 14.56^b$ & $49.79 \pm 4.03$ & $45.28 \pm 1.36$ \\
0.005 & 70 & $68.29 \pm 6.20$ & $69.55 \pm 4.59$ & $71.69 \pm 1.56$ \\
0.005 & 90 & $92.64 \pm 17.07$ & $85.33 \pm 15.80$ & $82.41\pm3.97$ \\
0.01 & 10 & $10.97 \pm 2.41$ & $9.96 \pm 1.87$ & $6.67 \pm 0.28$ \\
0.01 & 30 & $32.45 \pm 3.83$ & $32.26 \pm 3.03$ & $31.77 \pm 0.54$ \\
0.01 & 50 & $69.13 \pm 11.98^b$ & $48.04 \pm 3.75$ & $46.49 \pm 1.01$ \\
0.01 & 70 & $70.54 \pm 10.76$ & $68.21 \pm 4.61$ & $66.55 \pm 1.76$ \\
0.01 & 90 & $93.15 \pm 14.82$ & $91.32 \pm 8.72$ & $92.38 \pm 1.23$ \\
0.05 & 10 & $93.44 \pm 38.04^b$ & $8.71 \pm 1.63$ & $8.96 \pm 0.24$ \\
0.05 & 30 & $27.54 \pm 3.12$ & $28.07 \pm 2.72$ & $29.46 \pm 0.44$ \\
0.05 & 50 & $44.37 \pm 3.94$ & $45.71 \pm 3.31$ & $48.81 \pm 0.62$ \\
0.05 & 70 & $64.57 \pm 5.08$ & $65.77 \pm 3.82$ & $66.29 \pm 1.68$ \\
0.05 & 90 & $86.95 \pm 12.51$ & $85.50 \pm 7.23$ & $85.18 \pm 3.59$ \\
0.1 & 10 & $9.12 \pm 1.79$ & $9.09 \pm 1.64$ & $8.31 \pm 0.23$ \\
0.1 & 30 & $28.07 \pm 3.15$ & $27.81 \pm 2.68$ & $24.93 \pm 0.47$ \\
0.1 & 50 & $47.68 \pm 4.15$ & $45.86 \pm 3.39$ & $44.54 \pm 0.79$ \\
0.1 & 70 & $67.80 \pm 4.29$ & $68.30 \pm 3.53$ & $69.08 \pm 0.88$ \\
0.1 & 90 & $86.73 \pm 11.28$ & $84.58 \pm 7.39$ & $88.19 \pm 0.95$ \\
0.5 & 10 & $10.14 \pm 1.15$ & $10.29 \pm 1.15$ & $9.82 \pm 0.11$ \\
0.5 & 30 & $29.54 \pm 1.81$ & $29.88 \pm 1.74$ & $31.17 \pm 0.18$ \\
0.5 & 50 & $50.10 \pm 2.13$ & $50.60 \pm 1.94$ & $52.01 \pm 0.22$ \\
0.5 & 70 & $70.15 \pm 2.33$ & $70.62 \pm 1.96$ & $72.41 \pm 0.24$ \\
0.5 & 90 & $89.92 \pm 14.99$ & $90.45 \pm 1.63$ & $84.32\pm0.68$ \\
1 & 10 & $10.04 \pm 0.95$ & $10.09 \pm 0.96$ & $9.92 \pm 0.08$ \\
1 & 30 & $29.81 \pm 1.47$ & $29.75 \pm 1.46$ & $29.54 \pm 0.13$ \\
1 & 50 & $49.77 \pm 1.61$ & $49.61 \pm 1.60$ & $48.87 \pm 0.14$ \\
1 & 70 & $69.83 \pm 1.47$ & $69.63 \pm 1.46$ & $69.23\pm0.18$ \\
1 & 90 & $89.72 \pm 1.02$ & $89.58 \pm 1.00$ & $80.49 \pm 7.01$ \\
\enddata
\tablenotetext{b}{Cases that show bimodal posteriors.}
\end{deluxetable}

The ${\mathbb N}=1$ case is conceptually similar to \cite{Anumarlapudi2023}'s GMM approach, as both methods use MCMC-based approaches to measure the desired parameters from the on-pulse and off-pulse regions directly. $\mathbb N=1$ estimates the expected value to within 5\% in $\sim97\%$ of cases, and within the estimated error in $\sim 85\%$ of cases. Although the latter value is much larger than $\mathbb N=3$, the error bars for $\mathbb N=1$ are much larger. GMM estimates the expected value to within 5\% and within the estimated uncertainty in 74\% of cases.

One of the main differences between them is that GMM considers $\mu_{\rm N}$ and $\sigma_{\rm N}$ as free parameters. 
GMM and $\mathbb N=1$ appear to be primarily consistent at accurately estimating $n_{\rm F}$ for $\mu_{\rm P}\geq0.1$. For $\mu_{\rm P}<0.1$, the two methods start to differ in their estimations. This implementation of GMM struggles to measure $n_{\rm F}\leq50\%$, showing bimodal posteriors and hence overestimating the values by a large margin. $\mathbb N=1$ is more uncertain at lower $\mu_{\rm P}$, but it is more consistently accurate than GMM, measuring the expected value within error $\sim 85\%$ of the time.

Similar to $\mathbb N=3$, GMM also experienced bimodal posteriors. This resulted in poor estimations from GMM at lower $n_{\rm F}$ are due to bimodal posteriors.  However, unlike $\mathbb N=3$, reprocessing this implementation of GMM did not result in converged results, and the bimodal posteriors persisted. For example, at $\mu_{\rm P}=0.001, n_{\rm F}=10\%$, the corner plot (see Figure \ref{fig:cornerplot example} in Appendix) shows two possible solutions at $\sim10\%$ and $\sim100\%$, with the latter being more probable. GMM reports the median of the sample, which is $\sim94\%$ in this case. There is a similar case for $\mu_{\rm P}=0.001, n_{\rm F}=30\%$, where there are two possible solutions at $\sim30\%$ and $\sim80\%$, where the reported value is $\sim75\%$. A similar case is seen at $\mu_{\rm P}=0.05, n_{\rm F}=50\%$. All cases are indicated in Table \ref{tab: simset} and Figure \ref{fig:nf_comp}. 
% Although the cause for the bimodal posterior in $\mathbb N=3$ and GMM, but not $\mathbb N=1$, is unclear, they are easily identifiable in cornerplots.

$\mathbb N=1$ and GMM are similar algorithms; therefore, the significance of the differences between their measurements are unexpected.
The primary differences in their approaches are the consideration of $\mu_{\rm N}$ and $\sigma_{\rm N}$ as free parameters and the MCMC samplers. 

Less free parameters may be the reason we see no bimodal posteriors in $\mathbb N=1$. As the noise parameters are fixed, there is no ambiguity in the on-pulse contribution; however, for GMM, very weak pulses can be mistaken for noise, and hence, the degeneracy leads to bimodal posteriors. 
In $\mathbb N=3$, where there are only three free parameters, the degeneracy could instead arise from fitting multiple Gaussians close to the noise floor. It is unclear why fitting multiple Gaussians in this manner would lead to a bimodal posterior and not a multimodal posterior. It is possible this may change with higher $\mathbb N$.

Gaussian mixture model uses the \texttt{emcee} package's Ensemble sampler, an affine invariant of the Metropolis-Hastening (M-H) method \cite{2013Foremna, 2010Goodman}; $\mathbb N$sum uses the pymc package and uses a Hamiltonian Monte Carlo (HMC) based No-U-turn sampler \cite{2014Hoffman}. Therefore, the different sampling methods could explain the different results obtained by $\mathbb N=1$ and GMM.

The M-H method relies on random walks to sample the parameter space and converge on a solution. The HMC method avoids random walks and samples the parameter space by taking informed steps based on the gradient and momentum of the sampler. The informed sampling typically allows HMC approaches to converge faster than random walk approaches, making them more efficient at higher dimensions.

\subsection{Nulling Fraction Measurements}\label{sec:nullfmeasurements}

${\mathbb N}=3$ estimates $n_{\rm F}$ values are reasonable based on a visual assessment of the pulse stacks. Although these measurements do not always match $\mathbb N=1$ and GMM within uncertainty, the measurements are within expectation, e.g. returning low $n_{\rm F}$ for weakly nulling pulsars. From the simulated data, we know to consider uncertainties from $\mathbb N=3$ to be a slight underestimate; if true, this method likely agrees with the others. 

Cases where ${\mathbb N}=3$ differed significantly from the other methods were for weaker pulsars that had S/N $\sim 10 - 20$ throughout their 80\,min observation. These pulsars (J1919$+$0021, J1559$-$4438 and J1901$-$0906) were reported to have low $n_{\rm F}$; however, $\mathbb N=1$ and GMM methods reported consistently higher nulling fractions, as high as $n_{\rm F}=80$\%. For these pulsars, $\mathbb N=3$ consistently measured lower $n_{\rm F}$ than the other methods, which did not directly agree with the literature values within error, but still indicated that the pulsars rarely null. Although $\mathbb N=3$ jitters around the expected the $n_{\rm F}$ (see Figure \ref{fig:nf_comp}), it does so by a smaller margin $(<5\%)$. Regardless, $\mathbb N=3$ estimates much lower $n_{\rm F}$ than $\mathbb N=1$ and GMM. Literature values using older methods can themselves be biased, as outlined by \cite{Kaplan2018}. We report the S/N or flux densities of past observations, which we use to compare our results, for the aforementioned pulsars as well as other details in \S \ref{sec:weakpulsars}.

Similar to our simulated comparison, we typically see a large agreement between ${\mathbb N}=1$ and GMM measurements, for cases where the pulse fluence distribution appears to be normally distributed. In cases where the fluence is log-normally distributed, both ${\mathbb N}$sum methods produce poor fits, as expected, and likely inaccurate measurements. Additionally, for some pulsars (J0452$-$3418 and J0026$-$1955), we find that the pulse fluences are distributed via two normal distributions according to the Akaike information criterion (AIC). In these scenarios, $\mathbb N$sum measurements were also affected, as it only assumes the pulse fluences follow a singular Gaussian distribution. From our simulated data, it was expected that GMM would have difficulty converging on results for weaker pulsars that are expected to have low $n_{\rm F}$. However, we find $\mathbb N=1$ and GMM to similarly overestimate $n_{\rm F}$ for weaker pulsars. The reasons for this are unclear.
We did not observe bimodal posteriors for GMM and $\mathbb N=3$ on our real pulsars.

Aside from the comparison between methods, we also compared our measured $n_{\rm F}$ values with previously reported values for our sample of pulsars. For this, we only used GMM as it an established method in the field. Although there are some exceptions, which are discussed later in Sections \ref{sec:gmm_discuss} and \ref{sec:individualpsrs}, we find that many of the nulling fraction measurements are consistent between studies at various frequencies. 

\subsubsection{Log-normally Distributed Pulse Fluences}\label{sec:lognormaldist}

As indicated by Table \ref{tab: nftable}, we observe a log-normal distribution in the pulse energies for J0630$-$2834, J0953$+$0755, J1136$+$1551, J1239$+$2453 and J1456$-$6843. In some cases, Gaussians are inadequate to properly model the true distributions, leading to inaccurate measurements. For $\mathbb N=1$ case, as shown by the top right panel in Figure \ref{fig: J1456-6843_example} it can be seen clearly that the Gaussian fit to the histograms overestimates and underestimates different components of the distributions, resulting in an overall overestimation of $n_{\rm F}$ compared to the historical measurements.

The Gaussian mixture model faces a similar issue; \cite{Anumarlapudi2023}'s implementation allows for the fitting of a Gaussian with an exponential tail. This accommodates pulsars with log-normal-like distributions. In Table \ref{tab: nftable}, we use the `Gaussian with an exponential tail' model where possible for these pulsars.

Some of these pulsars showed extreme levels of log-normality. Even when three random pulses were summed, they remained in their log-normal shape; hence, even the $\mathbb N=3$ cases struggled to fit the data properly. This includes pulsars such as J1136$+$1551, J1456$-$5843, J0630$-$2834 and J0953$+$0755, where the Gaussians do not fit the histogram (see middle right panel of Figure \ref{fig: J1456-6843_example}). However, J1239+2453 lost its log-normality when its pulses were summed, resulting in a Gaussian shape for $\mathcal F$. Therefore, $\mathbb N=3$ estimated a more reasonable value in comparison to $\mathbb N=1$ and GMM. See Section \ref{j1239_sec} for more details on the pulsar's measurements.

\begin{deluxetable}{lcccccc}
\tablecaption{The nulling fraction measurements (as percentages) from the $\mathbb N$sum and Gaussian mixture modelling approach for a variety of nulling pulsars. The last three columns contain nulling measurements from the literature, the frequency at which they were taken and the reference. A majority of these values were obtained via the Nulling Pulsars catalogue \citep[][ \url{http://www.ncra.tifr.res.in/~sushan/null/null.html}]{2019Konar}.}
\label{tab: nftable}
\tablehead{
\colhead{PSR} &  \colhead{$\mathbb N$sum ($\mathbb N=1$)}& \colhead{$\mathbb N$sum ($\mathbb N=3$)} & \colhead{GMM} & \colhead{Literature}
&\colhead{Frequency}& \colhead{References}\\
& \colhead{(\%)}&\colhead{(\%)} & \colhead{(\%)}& \colhead{(\%)}&\colhead{(MHz)}& 
}
\startdata
% J0630-2834 && 0.68 $\pm$ 0.67 & 0.03 $\pm$ 0.04&
J0630$-$2834$^{x}$ & $0.05 \pm 0.05$ & $0.77 \pm 0.09$ & $0.03 \pm 0.05$&$\leq 2$ &$400-500$& (1) \\
&&&&$\leq 0.3$&$600-700$& (2)\\
&&&&$13.6\pm1.9$&$300-400$& (3)\\
% J0742-2822 && 2.27 $\pm$ 1.86 & 0.11 $\pm$ 0.14 &
J0742$-$2822 & $0.16 \pm 0.16$ & $5.85 \pm1.01$ & $0.11 \pm 0.15$&$\leq0.25$ &$400-500$& (1) \\
&&&&$\leq0.2$& $600-700$&(2)\\
% J1430-6623 && 0.68 $\pm$ 0.49 & 0.65 $\pm$ 0.78 &
J1430$-$6623 & $0.88 \pm 0.78$ & $2.75 \pm 0.49$ & $0.65\pm 0.78$&$\leq 0.05$ &$600-700$& (2) \\
% J1453-6413 && 0.59 $\pm$ 0.56 & 0.69 $\pm$ 0.15 &
J1453$-$6413 & $0.69 \pm 0.15$ & $1.11 \pm 0.19$ & $0.69 \pm 0.15$&$-$ & $1300-1500$&(4) \\
J0953$+$0755$^{ x} $ &$37.13 \pm 0.78$ & $31.82\pm 3.38$ & $35.83\pm1.03$ 
% ($54.13 \pm 0.87$) 
& $\leq 5$& $300-400$ &(1)  \\
&&&&$\leq5$& $400-500$ & (5)\\
&&&&$5.9\pm6.1$& $2000-5000$& (6)\\
J1136$+$1551$^{ x} $ &$17.14 \pm 1.11$ & $20.39 \pm 1.61$ & $17.20\pm2.32^{\alpha}$ 
% ($17.38 \pm 4.46$)
& $15\pm2.50$& $400-500$&  (1) \\
&&&&$5.9\pm6.2$& $2000-5000$ & (6)\\
&&&&$\sim15$& $300-600$& (7)\\
&&&&$14.4$& $300-400$ & (8)\\
J1239$+$2453$^{x}$ &$57.24 \pm 1.27$ & $18.69 \pm 0.71$ & $67.98\pm1.34$
% ($67.98 \pm 1.36$)
& $6.0\pm2.5$& $400-500$ &(1) \\
&&&&$7.0\pm3.0$& $300-400$ &(9)\\
&&&&$4.0\pm1.0$& $600-700$& (9)\\
&&&&$2.0\pm0.9$& $2000-5000$ &(6)\\
J1901$-$0906$^{zy}$ & $76.41 \pm 8.75$ & $9.13 \pm 0.79$ & $69.25 \pm 13.63$ & $2.9$ & $300-400$ & (3) \\
&&&&$5.6\pm0.7$&$600-700$&(3)\\
&&&&$29\pm4$& $300-400$ & (9)\\
&&&&$30\pm1$& $400-500$ & (9)\\
J2046$-$0421$^{ y}$ & $5.44 \pm 3.98$ & $5.9 \pm 1.2$ & $4.59 \pm 4.10$ & $-$ & $300-1400$ & (9) \\
J1913$-$0440 & $0.07 \pm 0.06$ & $0.57 \pm 0.15$ & $0.05 \pm 0.07$ & $\leq0.5$ & $400-500$ & (1)\\
J1919+0021 & $30.09 \pm 15.35$ & $1.34 \pm 0.34$ & $29.31 \pm 20.82$ & $\leq0.1$ & $400-500$ & (10) \\
J0452$-$3418 &$33.43 \pm 2.16$ & $19.45 \pm 0.70$ & $33. 50\pm 7.67$ & $34.0\pm6.0^{\beta}$ & $100-200$& (11) \\
J1456$-$6843$^{x}$ & $39.40\pm0.81$ & $23.69 \pm 1.75$ & $4.76\pm1.58^{\alpha}$
% ($12.02 \pm 0.95$) 
& $\leq3.3$ & $600-700$ &(2) \\
J0837$-$4135 &$2.18 \pm 1.76$ & $4.59 \pm 0.39$ & $1.98 \pm 2.03$& $\leq 1.2$ & $600-700$ & (2) \\
&&&&$1.7\pm1.2$& $600-700$ & (12)\\
J1559$-$4438$^{z}$ &$81.57 \pm 8.25$ & $2.58 \pm 0.69$ & $82.69 \pm 20.21$  & $\leq 0.01$ & $600-700$ & (2)\\
&&&&$0.24$& $600-700$ & (3)\\
J1645$-$0317 &$0.03 \pm 0.03$ & $1.4 \pm 0.28$ & $0.02 \pm 0.03$ & $\leq0.5$ & $300-400$ & (1) \\
&&&&$\leq0.25$& $400-500$ & (5)\\
J0026$-$1955 & $39.04\pm 1.85$&$20.46\pm 0.68$& $59.63\pm4.82$ & $\sim78^{\beta}$ & $100-200$ & (13)\\
&&&&$\sim58$& $300-500$ & (14)\\
\enddata
\tablecomments{
References: (1) \cite{1976Ritchings}, (2) \cite{1992Biggs}, (3) \cite{Basu2017}, (4) \cite{2012Burke-Spolaor}, (5) \cite{1995Vivekanand}, (6) \cite{Wang2020}, (7) \cite{2007Bhat}, (8) \cite{2007Herfindal}, (9) \cite{2017Naidu}, (10) \cite{1986Rankin}, (11) \cite{Grover2024a}, (12) \cite{2012Gajjar}, (13) \cite{Mcsweeney2022}, (14) \cite{2023Janagal} }
\tablenotetext{x}{These pulsars are very bright and exhibit a log-normal distribution.}
\tablenotetext{y}{Have persistent RFI.}
\tablenotetext{z}{These pulsars are very weak.}
\tablenotetext{\alpha}{The values were obtained using the Gaussian model with an exponential tail in GMM.}
\tablenotetext{\beta}{This measurement was made with the same data that was used in this work.}
\end{deluxetable}

% \begin{figure}[h]
%     \centering
%     \includegraphics[width=\linewidth]{Nillfraction_comprisons.png}
%     \caption{Caption}
%     \label{fig:Nullfraction_comparison}
% \end{figure}

% \begin{itemize}
%     \item using Gaussian mixture model
%     \item identify nulling fraction
%     \item also use novel approach of summing three random pulses and oberving the distribution and using mcmc to figure out the nulling fraction
%     \item 

% \end{itemize}

\subsection{Nulling Quasiperiodicity: Classification and Analysis}\label{sec:quasip_analysis}

Quasiperiodic nulling has been observed by multiple authors \cite[][etc.]{1970Backer,Basu2017,Basu2020,Tedlia2024,Xu2024}; however, the classification of this phenomenon may differ each time. Typically, the quasiperiodicity in nulling is measured by performing a fast Fourier transform (FFT) of the pulse fluences over time and measuring the dominant frequency.  A longitude-resolved fluctuation spectrum (LRFS) can also be used to measure quasiperiodicity. LRFS is essentially an average of the FFTs taken vertically over the on-pulse region. As nulling quasiperiodicity is on the order of 10s of pulses, features corresponding to these periods are on the low-frequency part of the spectra. However, low-frequency features are also attributed to modulation due to sub-pulses drifting (this value is also referred to as $P_3$) and quasiperiodic amplitude modulation. 
% Hence, the origin of structures in the Fourier spectra may be incorrectly attributed. 
Therefore, conclusions drawn from a collection of data from the literature, as done in \cite{Grover2024a}, may be subject to biases or discrepancies due to the varying classification criteria and methodologies employed by different studies.

To classify quasiperiodic nulling, we first remove the effects of the beam throughout the observations to remove any introduced red noise. We then identify the on-pulse region and make an array of mean energy values for each pulse. We then convert all values to a binary format with an intensity threshold distinguishing 1s (bursts) and 0s (nulls). The threshold can be chosen arbitrarily; we tested a range of thresholds but settled on using the first standard deviation of the off-pulse region intensity. This process can be made less arbitrary in the future with better methodologies for classifying pulses.

Once a binary array of the on-pulse region is created, we perform an FFT, which is then smoothed by convolving it with a window function. We ensure the convolution results in an array of the same size as the original FFT. We produce the power spectrum and normalise it such that the mean of the spectrum is one. Specifically, we normalised the spectrum by setting the median to a natural log of two. We identify the peak in the Fourier spectrum and fit a Gaussian to the peak structure. If the peak structure is $>3\sigma_{\rm rms}$, where $\sigma_{\rm rms}$ is the root-mean-square of the high-frequency component of the spectrum, the spectrum can be classified to show quasiperiodicity. Although this method would disregard high-frequency quasiperiodicity, studies show that quasiperiodic pulsars have nulling periods $\geq10P$ \citep{Wang2007,Basu2020}.

To measure the periodicity, we fit a Gaussian around the target structure and use the mean of the Gaussian as well as the covariance of the fit as the measurement and uncertainty.

In our data, we identify four pulsars showing nulling quasiperiodicity according to our criteria, shown in Table \ref{tab: qptable}. We also measure the quasiperiodicity of a fifth pulsar, which doesn't meet our criteria, but it has been reported to show quasiperiodicity in the literature. We measure nulling quasiperiodicity in three pulsars, J1453$-$6413, J0950$+$0755 and J0026$-$1955, that was not previously reported or measured. Two pulsars, J1453$-$6413 and J0950$+$0755, have short periods, sitting among the fastest pulsars to show quasiperiodicity, while J0026$-$1955 appears much older (see Figure \ref{fig:ppdot}).

For the rest of the pulsars in our sample, we observe no significant quasiperiodicity.

\begin{deluxetable}{lcc}
\tablecaption{Nulling quasiperiodicity measurements which have not been previously classified to show periodic nulling. Nulling quasiperiods are measured in pulses.}
\label{tab: qptable}
\tablehead{
\colhead{PSR} & \colhead{Nulling Quasiperiod } & \colhead{Literature Measurements} 
}
\startdata
  J1453$-$6413&$117.93\pm0.87$&$-$
  \\
  J0953+0755&$61.08\pm0.81$&$-$
  \\
   J0026$-$1955&$136.66\pm9.97$& $-$
  \\
  J1136+1551&$29.94\pm0.34$&$29\pm2$
  \\
  J0452$-$3418&$36.84\pm0.29$&$42^{+1.5}_{-1.3}$
  \\
  J1239+2453 & $\sim45/24^{x} $&$26\pm 5$
  \\
\enddata
% \tablecomments{X}
\tablenotetext{x}{The periodic feature in this pulsar's spectra did not meet our criteria; however, the structure appeared noteworthy.}
% \tablenotetext{y}{Have presistent RFI.}
\end{deluxetable}

\subsection{Individual Pulsars:}\label{sec:individualpsrs}
\subsubsection{Weak pulsars} \label{sec:weakpulsars}
J1559$-$4438, J1901$-$0906 and J1919$+$0021 are all extremely weak pulsars in our observations with S/N $\sim 10-20$ in the 80 minutes observed using the MWA.

\paragraph{J1559$-$4438 (B1556$-$44)}
J1559$-$4438
% is extremely weak at $\sim 155$MHz with a S/N $\sim 10$ in 80 minutes of data. Given that the pulsar 
has a flux density of 120\,mJy at 300\,MHz \citep{2005Manchester}\footnote{https://www.atnf.csiro.au/research/pulsar/psrcat/}, it must have experienced a turnover for the observation to be this weak. The pulsar also reaches a maximum beam power of 60\%, indicating that the low S/N is due to the intrinsic flux of the pulsar. Due to this, all three algorithms have overestimated the nulling fraction $>80\%$, whereas past measurements have been $<1\%$. \cite{1992Biggs} observes all $7300$ individual pulses in their observation, justifying their upper limit of $n_{\rm F}\leq0.01\%$. 

% J1919$+$0021 has a similar scenario with a S/N $\sim13$ in a full SMART observation. The 

\paragraph{J1901$-$0906}
Other than a low S/N, the J1901$-$0906 observation also had significant RFI. This pulsar is intrinsically weaker at our frequencies, having already experienced a turnover given the flux at 350\,MHz and 400\,MHz is 8.3\,mJy and 11\,mJy, respectively \citep{2005Manchester}. The $n_{\rm F}$ measurements are also overestimated at $n_{\rm F}\geq 60\%$ with past measurements also showing $n_{\rm F}\leq30\%$. The measurements by \cite{2017Naidu} at 325 MHz and 610 MHz have S/N per pulse of 1.3 and 1.6.

\paragraph{J1919$+$0021 (B1917$+$00)}
For J1919$+$0021, historical measurements indicate $n_{\rm F} \leq 0.1\%$ \citep{1986Rankin}, whereas our measurements indicate $n_{\rm F}\sim30\%$ with GMM and $\mathbb N=1$, and $1.34\pm0.34\%$ by $\mathbb N=3$. Although the pulsar is a steep spectrum source, expected to be at $\sim 92$\,mJy at 150\,MHz \citep{2005Manchester}, from the EPN database of pulsar profiles\footnote{https://psrweb.jb.man.ac.uk/epndb/}, the pulsar has an S/N$ \sim 10$ at $\sim135$\,MHz. However, this value should be taken with caution as the corresponding cited papers on the EPN database do not show the profile or mention details about its flux. In our observations, the pulsar is only observed at $20\%-30\%$  of the maximum beam power, but as suggested by the EPN database, a spectral turnover is likely the cause for the low S/N in our observations. The value measured in the literature was taken with the Arecibo telescope between 400-500\,MHz in a preprint by Nowakowski and Hankins \citep{1986Rankin}. Although the S/N of their observations is unknown, \cite{1981Rankin} reported a flux density of $\sim30$\,mJy using the same telescope. Given the brightness of the pulsar in their observations, it is likely their $n_{\rm F}\leq1$ estimate is accurate. 
% As SMART observations are drift scans, the beam power towards these pulsar ranges from  0.2 - 0.3 of the full beam power. Although the variation is minimal, the pulsar was already a weak source for the MWA. Hence, the variation in the beam may have affected its visibility and, therefore, made $n_{\rm F}$ appear larger.

\begin{figure}[h]
    \centering

    % \caption{J1453-6413}
    \includegraphics[width=0.48\linewidth]{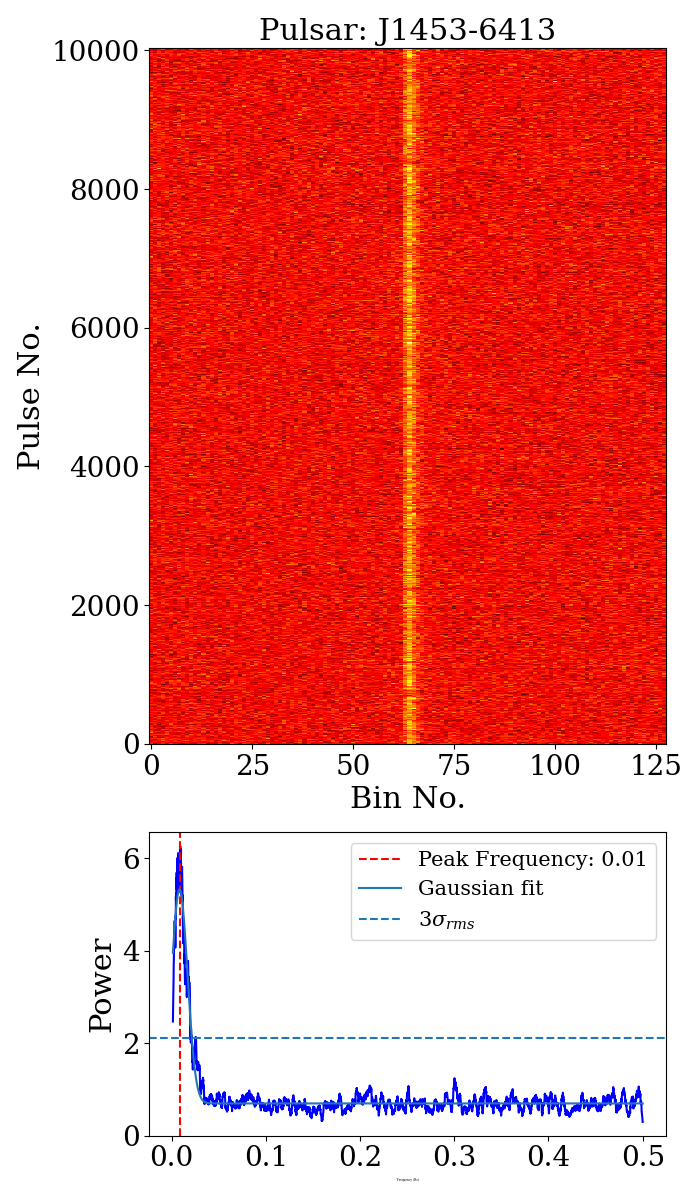}
    % \caption{J0953+0755}
    \includegraphics[width=0.48\linewidth]{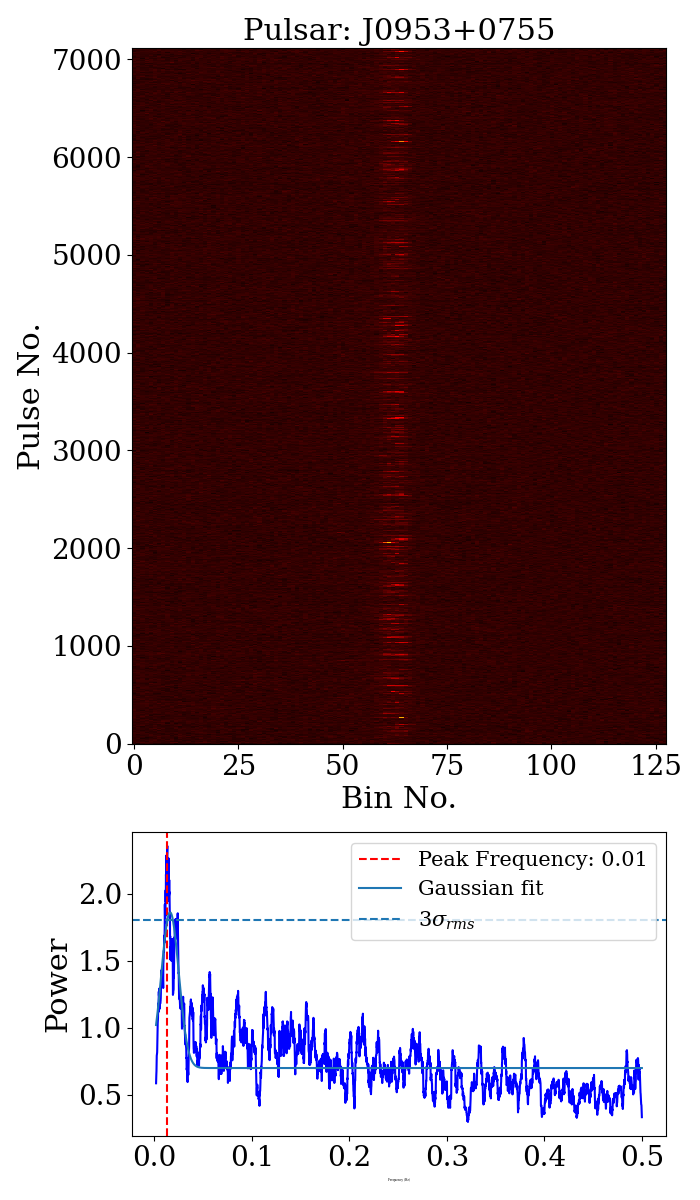}
    \caption{The top plots show pulse stacks of pulsars J1453$-$6413 and J0953$+$0755, and the bottom plots show the FFTs of the pulse stacks. The peak in the FFTs is due to modulation in the pulse stacks. The red dashed vertical line indicates the peak, and the blue dashed horizontal line indicates the 3$\sigma_{\rm rms}$.}
    \label{fig:both}
\end{figure}

\subsubsection{J0026$-$1955}\label{sec:j0026}

This pulsar is a highly nulling, sub-pulse drifting pulsar which was rediscovered in the SMART survey \citep{Mcsweeney2022}. This pulsar has multiple sub-pulse drifting modes and has been observed to have an evolving drift rate within drift sequences \citep{2023Janagal}. This pulsar also continues its drift pattern from where it left off after a nulling sequence, which is also known as `memory across nulls' \citep{1982Filippenko}. 

To measure this pulsar's nulling fraction, we used the same observation as \cite{Mcsweeney2022}, where they used a threshold method to measure $n_{\rm F} \sim78\%$. Using GMM, we find a much lower value of $59.63\pm4.82 \%$. We used GMM with 2+1 Gaussian components (where the +1 refers to the nulling component), as suggested by the AIC values, rather than 1+1; this suggests this pulsar has multiple fluence distributions. This is likely why the other $\mathbb N$sum measurements are different and likely incorrect, as they assume a single Gaussian distribution for the fluence. The GMM measurement of $59.63\pm4.82 \%$ is given confidence with \cite{2023Janagal}'s measurement of $\sim58\%$ using GMRT observations.

In \cite{Mcsweeney2022}, the LRFS of this pulsar did not show structure indicating nulling quasiperiodicity, unlike our analysis. This may be due to the methodology of our analysis, where all pulse phase structure is combined into one value. However, \cite{Mcsweeney2022} used multiple observations, whereas we only used observation B07; therefore, it is possible our analysis showed non-persistent quasiperiodicity. 

\subsubsection{J0452$-$3418}\label{sec:j0452}

Similar to J0026$-$1955, J0452$-$3418 is also a pulsar discovered using the SMART survey with the MWA \citep{Grover2024a}. We use an observation from the discovery paper to measure $n_{\rm F}$; \cite{Grover2024a} also used GMM to measure the nulling fraction to be $35\pm8\%$. They fit 1+1 Gaussians to their data. We used GMM with 2+1 Gaussians, as suggested by the AIC measurements, and hence measure a slightly lower value of $33.50\pm7.67\%$. We also measured the on-pulse region systematically by fitting Gaussians to the integrated profile and used the full-width at 10\% of the maximum power as the cut-offs for the on-pulse region. $\mathbb N=1$ method, which fits 1+1 Gaussians, also measured $n_{\rm F}=33.43\pm2.16\%$; however, the Gaussian fit to the histogram in this case is visibly worse. $\mathbb N=3$ measured a much lower value of $19.45\pm0.70\%$, likely due to the assumption of the on-pulse fluence being drawn from a single Gaussian distribution rather than two.

In our previous paper, we measured the nulling quasiperiodicity of J0452$-$3418 using three individual observations \cite{Grover2024a}. In this study, we only use one observation, hence the discrepancy between the measured and literature values in Table \ref{tab: qptable}.

\subsubsection{J0953$+$0755 (B0950$+$08)}
J0953$+$0755 is a well-studied pulsar, being one of the first pulsar discoveries after CP 1919 due to its brightness and proximity. It has been observed to have large pulse amplitude variability \citep{1973Smith}, emit giant pulses \citep{2004Cairns}, and null \citep{1974Hesse}. 

These features mean that the fluence distribution is extremely log-normal and has an extended tail, which accounts for its extremely bright pulses. These factors affected our $\mathbb N$sum measurements as we assume a Gaussian pulse fluence. Additionally, we were also unable to fit the Gaussian with exponential tail model with GMM and obtained inaccurate $n_{\rm F}$ measurements for all methods.

All of the pulse variability features can contribute to a quasiperiodicity measurement via the Fourier spectrum. The spectrum shows multiple sharp, small-amplitude features with one dominating feature corresponding to a quasiperiodicity of $61.08\pm0.81\,P$. 

\subsubsection{J1136$+$1551 (B1133+16)}
J1136$+$1551 is a very bright pulsar showing log-normally distributed pulses. Although $\mathbb N=1$ and GMM $n_{\rm F}$ measurements for this pulsar appear to be very similar at $n_{\rm F}\sim 17\%$, the $\mathbb N=1$ fits are poor. Using GMM with an exponential tail model, we coincidentally measure a similar $n_{\rm F}\sim 17\%$.  This agrees well with the multiple past measurements. Although $\mathbb N=1$ it measures the same result, it is not reliable for such cases.

J1136$+$1551 is also among the first pulsars reported to show quasiperiodic nulling \cite{2007Herfindal}. Our measurements agree well with prior nulling quasiperiodicity measures of $\sim 29$ $P$ \citep{Basu2017}. 

\subsubsection{J1239$+$2453 (B1237+25)}\label{j1239_sec}

J1239+2453's pulse fluences appear to be distributed log-normally; however, the $\mathcal F$ distribution appears more Gaussian. Therefore, $\mathbb N=1$ and GMM (with the Gaussian only model) measure $n_{\rm F}$ to be an order of magnitude ($60\%-70\%$) higher than the multiple literature values. In contrast, $\mathbb N=3$ measurement is much lower ($\sim 19\%$), and much closer to the other measurements.  A closer look at the histogram shows log-normal traits, which is why the Gaussians do not perfectly match the histogram. However, this pulsar is a good example of the borderline case where pulse fluences begin to appear more Gaussian the weaker their observations are. Hence, for weaker pulsars, the `Gaussian fluence' assumption may still lead to reasonable $n_{\rm F}$ values.

In terms of quasiperiodicity, J1239+2453 does not meet our criteria of a strong enough peak. It does show two structures, which could be indicative of a quasiperiodic behaviour. The pulsar has previously measured nulling quasiperiodicity of $\sim$26 pulses \citep{Basu2020}. We find a primary peak and a secondary peak, which suggest a quasiperiod of $\sim 43$ pulses and $\sim 24$ pulses, respectively. \cite{Herfindal2009DeepBehaviour} suggested that the quasiperiodicity may be due to core-active and absent submodes rather than nulling.

% \begin{itemize}
%     \item perform fft and 2dfs to measure sub-pulse drifting features and nulling quasiperiodicity
% \end{itemize}
\subsubsection{J1453$-$6413 (B1449$-$64)}

J1453$-$6413 is a southern-sky pulsar that has been known to null; however, not many studies focused on the single-pulse analysis of this pulsar, let alone observing it at low frequencies of $\sim 155$\,MHz. \cite{2012Burke-Spolaor} identified the pulsar nulls, but they provided no nulling fraction; they only reported the detection of 1828/3107 pulses. At first glance, the pulsar often appears to be nulling; however, further inspection shows a mix of amplitude modulation and nulling. Nulling here refers to the pulses having an arbitrary unit of flux $<0$, hence, it is possible that the pulsar simply modulates in amplitude, changing to a weaker emission mode.

We also find the pulsar to gradually decrease in amplitude before it switches to the weaker mode and or nulls, similar to what is observed by \cite{1986Deich}. Nulling pulsars are known to sharply change amplitude between emitting and nulling. Hence, a gradual change in amplitude before a null is uncommon.

Regardless of the origin of the change in amplitude, our Fourier analysis shows strong quasiperiodicity ($117.93\pm0.87\,P$) in the behaviour of the pulsar that has not been previously reported.

\subsubsection{J2046$-$0421}
J2046$-$0421 (B2043-04) observations contain a significant amount of RFI that affected a large number of individual pulses; however, the pulsar is bright enough to be visually distinguishable from the RFI. Although no prior $n_{\rm F}$ measurements have been made for J2046$-$0421, ${\mathbb N}=1$, $\mathbb N=3$ and GMM all agree on a value of $\sim5\%$. However, based on prior reports on how little the pulsar nulls \citep{2017Naidu}, the true $n_{\rm F}$ is likely smaller.

\section{Discussion}\label{sec:discusison}
\subsection{$\mathbb{N}$sum Method}

Nulling pulses and weak pulses can often be mistaken for one another based on the noise floor of the observing telescope. This generally leads to an overestimation of nulling parameters using current methods, as noted by \cite{Kaplan2018}. To overcome the issue of misidentification, we attempted to extract extra information from pulse stacks by summing two or more random pulses together to create a histogram of summed fluence. For low S/N pulsars, the summed fluence distribution becomes distinguishable from the single pulse distribution by virtue of its higher energy tail, which corresponds to multiple ``on'' pulses being added together. This effect naturally increases as more pulses are co-added (larger $\mathbb N$). 

Based on simulated comparisons in Table \ref{tab: simset}, $\mathbb N=3$ jitters around the expected $n_{\rm F}$ for a majority of the cases. This behaviour can be generalised as the method having a lack of precision in estimating $n_{\rm F}$. This lack of precision could be attributed to the complexity of the method and limitations on the MCMC samples. Considering a simple case where pulse fluences are distributed via a simple Gaussian, $\mathbb N=1$ and GMM have to deconstruct the mixture into two Gaussians, whereas for $\mathbb N\geq2$ cases, the number of Gaussians to be extracted are $\geq3$. Additionally, for weaker pulsars, $\mathbb N\geq2$ methodologies extract multiple Gaussians from a seemingly singular Gaussian distribution as shown by the middle right panel of Figure \ref{fig: J1919+0021_example}, which likely adds some level of degeneracy in the parameters of the extracted Gaussians.

Although the method shows some inaccuracies, it typically estimates the $n_{\rm F}$ within $<5\%$ of the expected value. Uncertainties reported by $\mathbb N$sum are simply the standard deviation of the samples in the MCMC process, which may not always reflect the uncertainty in $n_{\rm F}$. To combat the inaccuracies mentioned above, we suggest an uncertainty of $5\%$ should be used.
% Hence, using an uncertainty of $\pm5\%$ will truly reflect the range of possible nulling fractions as suggested by the simulated data.

In the simulated data, $\mathbb N=3$ also occasionally shows a lack of convergence and bimodal posteriors. While a lack of convergence can generally be addressed by increasing the number of steps taken by the MCMC samples, the bimodal posterior highlights the possibility of two solutions. 
% This issue also occurs with the GMM approach; however, the issue persists with multiple iterations of MCMC. 
For an iteration of $\mathbb N=3$ at $\mu_{\rm P}=0.5,n_{\rm F}=90\%$, the lower $n_{\rm F}$ solution at $\sim0\%$ corresponds to the parameters $\mu_{\rm P}\sim0.05,\sigma_{\rm P}\sim0.35$, whereas the solution $n_{\rm F}\sim 90\%$ returns  $\mu_{\rm P}\sim0.6,\sigma_{\rm P}\sim0.05$. That is, the possible solutions are pulses with very weak and broadly distributed fluences with low nulling fractions or brighter, sharply distributed fluences with higher nulling fractions.  Given the relation of the parameters used to minimise the likelihood, bimodal posteriors are unsurprising and are resolved by reprocessing the MCMC. 

% This issue may be resolved by also reporting the mode as well as the mean of the MCMC samples for cases of bimodal posteriors, leaving the choice of the appropriate parameter to the user.

When comparing this performance on real pulsars, we find that this method shows comparable measurements for pulsars with Gaussian pulse fluences. The method does not find any bimodal posteriors for real data. For weaker pulsars in our observations, this method measures $n_{\rm F}$s closer to the literature values, which are taken with observations of higher S/N. 

% We find some pulsars to be very weak in our observations, and $\mathbb N=1$ and GMM estimate the nulling fractions to be very high (see Section \ref{sec:weakpulsars}). However, past observations using more sensitive telescopes report $n_{\rm F}$ to be very small for these pulsars, which is correctly estimated by $\mathbb N=3$. 

For $\mathbb N=1$, the simulated data indicate that it performs consistently well throughout all $\mu_{\rm P}$. Even at low $\mu_{\rm P}<0.01$, the method shows appropriate measurements. The estimated $n_{\rm F}$ from $ \mathbb N=1$ are very comparable with GMM measurements for higher $\mu_{\rm P}>0.01$, which is expected as the two have similar algorithms with different MCMC approaches and two extra parameters. However, for real pulsars with low S/N, this method weakly converges on $n_{\rm F}$ values differing strongly from the historical values, similar to GMM. This is unexpected, given the method's performance on the simulated data. 

In the case where the historical values are closer to the true nulling fraction, possible reasons for these poor estimates by GMM and $\mathbb N=1$ are unclear. When comparing the estimated $\mu_{\rm P}$ and $\sigma_{\rm P}$ between $\mathbb N=1 \text{ and } \mathbb N=3$, we find $\mathbb N=1$ estimated fluences to be weaker and narrowly distributed for these pulsars, whereas $\mathbb N=3$ estimated brighter fluences to be more broadly distributed. For example, in the case of J1919$+$0021, $\mathbb N=3$ estimated $\mu_{\rm P}\sim0.27, \sigma_{\rm P}\sim 1$ whereas $\mathbb N=1$ estimated $\mu_{\rm P}\sim0.03, \sigma_{\rm P}\sim 0.08$. As there have not been any studies on the fluence distributions of these pulsars, it is difficult to compare which is truer. It is possible that the historical values are incorrect or the pulsar shows a strong change in $n_{\rm F}$ over time; however, this is unclear due to the current lack of $n_{\rm F}$ measurements for these pulsars.

As we have observed the difference in results between $\mathbb N=1$ and $\mathbb N=3$, it raises the question of numerical performance at higher $\mathbb N$. Ideally, larger $\mathbb N$ better identify emission in weak pulsars as noted in Section \ref{sec:classnull}. However, based on the results in Section \ref{sec: comparison}, $\mathbb N=1$ and $\mathbb N=3$ show comparable results for low $\mu_{\rm P}$. How these results propagate further with larger $\mathbb N$ has not yet been explored. At very high $\mathbb N$ (e.g. $>10$), $\mathcal F^{\mathbb N>10}$ becomes very wide, and how this affects the MCMC analysis is not intuitively obvious. It is possible that some degeneracy is introduced; however, the exact effects are yet to be studied.

There are clear examples where the $\mathbb N$sum approach fails to work. For pulse fluence distributions that are log-normally distributed, the Gaussian-based method is not appropriate, as mentioned in Section \ref{sec:lognormaldist}. Additionally, for cases where the fluences are described by multiple distributions, such is the case with J0452$-$3418 (Section \ref{sec:j0452}) and J0026$-$1955 \ref{sec:j0026}, the current version of $\mathbb N$sum is not appropriate, as it expects fluences to follow a single Gaussian distribution. Although the corner plots may show good convergence, the Gaussians do not visually match the histogram. These poor fits often result in an out-of-place spike in the 3-nulling-pulse case that clearly does not match the data, as seen in Figure \ref{fig: J1456-6843_example}. Although for bright pulsars with fluences following log-normal or multi-Gaussian distributions that rarely null, $\mathbb N$sum may still indicate a low $n_{\rm F}$, it is generally not effective for fluences for such distributions.

In an attempt to overcome this issue, we implemented a log-normal distribution for the pulse fluence and a Gaussian for the nulling component. However, typical mixture modelling algorithms will not fit a log-normal distribution to data with negative values, as was the case with our data. Instead, we removed all negative values of the data and attempted to fit a half-normal distribution and a log-normal distribution, assuming the noise had a zero mean. However, this also fails to fit the data adequately. Future improvements of this algorithm could attempt to overcome this issue by implementing `Gaussian with an exponential tail' distribution as \cite{Anumarlapudi2023} did with GMM. This would be equivalent to convolving a lognormal distribution (for the emission) with a Gaussian distribution (to model the noise that gets mixed with it); however, there is currently no closed-form solution for this \citep{FURMAN2020}.

%As we only measure $n_{\rm F}$ with our code, 
We also attempted to measure the nulling duration parameter. 
%We also attempted to this 
Similar to the $\mathbb N$sum method, which used random pulses, we tried using sums of consecutive pulses in order to fit for the mean nulling duration and mean burst duration rather than a nulling fraction directly. This was attempted by assuming the nulling duration and burst durations are Poisson processes, and hence the duration between one bursting event and another bursting event can be described by the exponential equation $\lambda e^{-\lambda t }$ with $1/\lambda$ being the mean burst duration, vice versa for the nulling duration. The transition between nulling and bursting can be assumed with a Kolmogorov forward equation \citep{1949Fuller}. Along with the existing parameters, we also solve for additional parameters that describe the transition, $\lambda$ and hence the burst and null durations.
In this scenario, $\mathbb N$ would dictate the sensitivity to certain nulling and burst durations. The described methodology could be adapted to robustly measure null and burst durations.

\subsection{Gaussian Mixture Model}\label{sec:gmm_discuss}

The Gaussian mixture model had only been tested in \cite{Kaplan2018} for $\mu_{\rm P}=5$, whereas we test it for even weaker energies ($\mu=0.001$). We find that for simulated pulses of low mean fluence, the method shows bimodal posteriors for $n_{\rm F}$, even with multiple iterations of MCMC. As suggested previously, this likely due to $\mu_{\rm N}$ and $\sigma_{\rm N}$ being free parameters. Unlike $\mathbb N=3$, bimodal posteriors are only present for the standard deviation of the pulse fluence and the nulling fraction, while the mean fluence is the same for both solutions. Additionally, for GMM, the expected solution was the least probable of the two solutions. For example, for the simulated parameters $\mu_{\rm P}=0.001, n_{\rm F}=10\%$, the possible solutions are $\sim 10\%$ and $\sim 100\%$ where the latter is much more probable, as seen in Figure \ref{fig:cornerplot example} of Appendix. The bimodal posteriors may be resolved by keeping the noise parameters fixed or by simply adding $\mu_{\rm P}> \mu_{\rm N}$ as a prior.

The bimodal posteriors do not appear for real data. Instead, for weaker pulsars, e.g. J1919$+$0021, GMM fails to converge strongly on a solution, reporting the median value of $n_{\rm F}\sim 30\%$. Similar to $\mathbb N=1$, GMM also estimates a smaller mean $\sim 0.03$ and a smaller standard deviation $\sim 0.07$, in comparison to $\mathbb N=3$. Similar to $\mathbb N=1$, without more studies about the fluence distribution of the weak pulsars in our sample and more $n_{\rm F}$ measurements, it is difficult to consider arguments favouring GMM and $\mathbb N=1$ measurements over $\mathbb N=3$. 

% GMM as an approach is generally flexible. Unlike $\mathbb N$sum where using different fluence distributions can be non-trivial, the distributions in GMM can be adjusted as demonstrated by \cite{Anumarlapudi2023}. 

\subsection{Revisiting the Quasiperiodically Nulling Population in the $P \dot P$ Diagram}

\cite{Grover2024a} showed that quasiperiodically nulling pulsars in the literature sit primarily within the death valley in the $P\dot P$ diagram. A similar study of the quasiperiodically nulling population and their correlation with pulsar parameters, specifically, $\dot E$ was performed by \cite{Basu2020}. Whereas they do not test for a correlation between quasiperiodicity and period, they note that a majority of their sample had rotation periods $P>0.1$\,s, i.e. longer than the average pulsar. However, this is likely an observational bias as nulling has not been observed in short-period pulsars (millisecond pulsars). One of the main reasons for this is due to their low S/N per pulse, which is insufficient in currently available observational data.

We expand their sample by measuring nulling quasiperiodicity in previously unmeasured pulsars, by adding newer and older literature measurements that were absent in \cite{Basu2020, Grover2024a}'s analysis, and including nulling pulsars from \cite{Song2023ThePulsars} that have a $P_3$ measurement but show no sub-pulse drifting.

\begin{figure}
    \centering
    \includegraphics[width=0.7\linewidth]{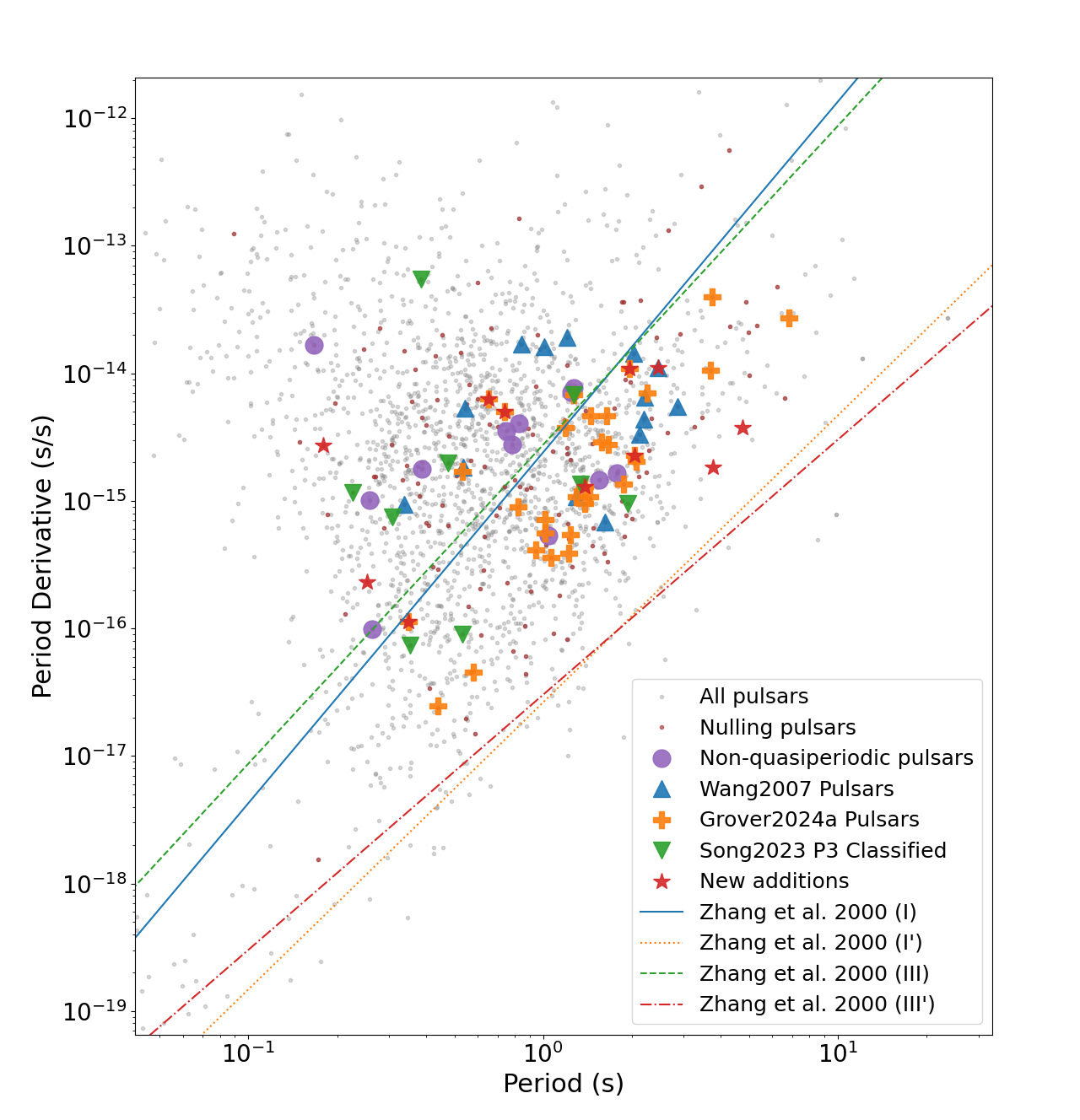}
    \caption{An updated $P \dot P$ from \cite{Grover2024a} which shows quasiperiodically nulling pulsars from multiple studies. We include measurements from multiple studies, as shown by the legend. For details on the studies, see the text. }
    \label{fig:ppdot}
\end{figure}

In Figure \ref{fig:ppdot}, we plot a $P\dot P$ diagram with the updated population of quasiperiodically nulling pulsars. We find the population is more sparsely spread compared to \cite{Grover2024a}'s sample. In red stars, we have the pulsars identified to show quasiperiodically nulling or have updated values from \cite{Tedlia2024, Wang2020, 2021Wang, 2023Wang, Xu2024} and this paper. In blue triangles, we plot pulsars identified from \cite{Wang2007} to show quasiperiodically nulling. A majority of this sample had very large quasiperiods with a median of $\sim 150\,P$; in contrast, the list of quasiperiods reported in \cite{Basu2020} had a median of $\sim 50\,P$. These larger quasiperiods could be attributed to \cite{Wang2007}'s methodology, as they summed up to 10\,s of data at a time to obtain larger S/N. The pulsars they sample also have a much lower declination in comparison to \cite{Basu2020}, therefore, they could be sampling a different population.

\cite{Song2023ThePulsars}'s paper analysed the sub-pulse modulation of $\sim 1200$ pulsars. For some pulsars, they measured a periodicity, which they labelled as a `$P_3$' value, but the pulsars did not appear to show signs of sub-pulse drifting. Instead, some pulsars in this group were reported to show nulling elsewhere in the literature. We suggest that the 
periodicity reported by \cite{Song2023ThePulsars} for this subset is due to quasiperiodic nulling. Therefore, we have added this group to the $P\dot P$ diagram as green triangles. As we discuss in the following Section, the position of the majority of this sample in Figure \ref{fig:edotplot} (the red triangles) suggests that the observed periodicity is not due to nulling.

Additionally, for comparison, we plot pulsars from our sample that do not show quasiperiodic nulling. We find the mean of this sample sitting on the left-hand side of the plot, outside the death valley. Although this is only a sample of 12 pulsars, thus far, they support a connection between quasiperiodically nulling pulsars and death lines.

\subsection{Nulling Quasiperiodicity and $\dot E$}
\cite{Basu2017,Basu2020} plotted the period of modulation for a phenomenon (e.g. $P_3$, quasiperiodic nulling, amplitude modulation and indeterminate periodic modulation) and the spin-down energy loss ($\dot E$). In \cite{Basu2020}'s Figure 4, they show that pulsars with different types of modulation lie in different parts of the figure. Modulations caused by sub-pulse drifting last typically for $1P \sim10P$ and those pulsars have a measured $\dot E$ of $10^{30} \,{\rm erg\,s^{-1}}\sim 10^{32}\, {\rm erg\,s^{-1}}$. Quasiperiodically nulling pulsars show modulations for $10P\sim100P$ with $\dot E$ ranging from $10^{31}\, {\rm erg\,s^{-1}}\sim 10^{33}\, {\rm erg\,s^{-1}}$. Other types of modulation related to mode changing or other effects lie in a similar period range to quasiperiodic nulling but the $\dot E$ range extends from $10^{30}\, {\rm erg\,s^{-1}}\sim10^{34}\, {\rm erg\,s^{-1}}$. 

\begin{figure}
    \centering
    \includegraphics[width=\linewidth]{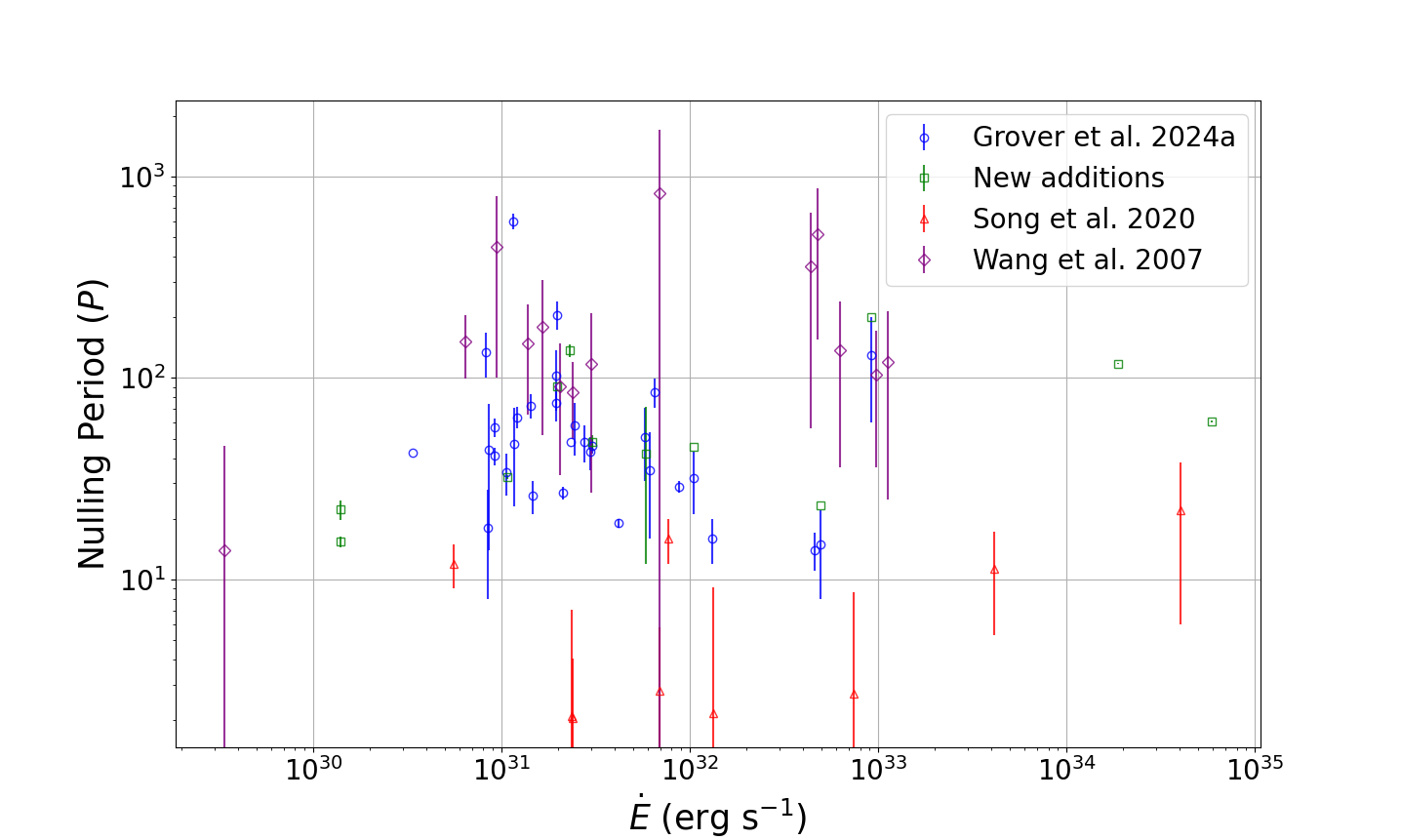}
    \caption{The nulling period of pulsars plotted against their spin-down luminosity. The different colours and symbols represent sources from different papers as per the legend. The new additions specifically include pulsars measured to show nulling quasiperiodicity from this work, \citet{Wang2020,2021Wang, 2023Wang,Tedlia2024,Xu2024}.}
    \label{fig:edotplot}
\end{figure}

%merge 
Following on from their work, we recreated the figure, plotting only quasiperiodically nulling pulsars. As seen in Figure \ref{fig:edotplot}, we expand their sample of quasiperiodically nulling pulsars by using pulsars listed in \cite{Grover2024a}, \cite{Wang2007}, quasiperiodically nulling candidates in \cite{Song2023ThePulsars}, as well as other measurements by \cite{2021Wang, 2023Wang, Xu2024, Tedlia2024} and this work. We find pulsars listed in \cite{Grover2024a} and \cite{Wang2007} lie in two islands, one between $10^{31} - 10^{32} {\rm  erg  \,s^{-1}}$ and one at $10^{33} {\rm erg \,s^{-1}}$, which build upon the regions present in \cite{Basu2020}. Some pulsars taken from \cite{Song2023ThePulsars}, which we expected to show quasiperodic nulling, lie well below the island. This departure from the main group could suggest a different origin of the measured quasi-periodicity. Based on the magnitude of the periodicity ($<10P$), the modulation observed in these pulsars matches that of sub-pulse drifting pulsars in \cite{Basu2020}'s Figure 4. However, the five pulsars that lie within that region have a larger average $\dot E$, but they continue to follow the anti-correlation trend found in \cite{Basu2017,Basu2020}.

We also note that the new additions also lie primarily on the main islands. Two pulsars that lie far off to the right are J1453$-$6413 and J0953$-$0755, for which we have measured quasi-periodicity. Based on their positions in Figure \ref{fig:edotplot}, these lie in the same region as pulsars identified to show mode changing or simple amplitude modulation. The single-pulse data of these pulsars do show amplitude modulation as well as nulling. It is possible that the amplitude modulation may have a stronger periodicity in these objects than the nulling, hence their positions in Figure \ref{fig:edotplot}.
%merge 
However, if J1453$-$6413's and J0953$-$0755's periodicity is shown to originate purely from amplitude modulation by a more sensitive telescope, then it would validate the use of Figure \ref{fig:edotplot} to classify origins of quasiperiodicities. Additionally, limiting quasiperiodic nulling to $\dot E<10^{33}{\rm erg s^{-1}}$ may imply that quasiperiodicity is an evolutionary step, at least nulling pulsars, in the later stages of their life cycle.

Alternatively, if the quasiperiodicity is due to nulling, then their position in Figure \ref{fig:edotplot} implies a likeness between the origins of amplitude modulation and quasiperiodic nulling, as mentioned in \cite{Basu2020}. Nulling has been suggested to be a weaker, undetectable mode of emission rather than a cessation of emission $-$ pulsars such as J1453$-$6413, which appear to show amplitude modulation and nulling, could be the key to this conjecture. 

It should be noted that \cite{Basu2017, Basu2020, Grover2024a} and this work draw conclusions from measurements and classifications of quasiperiodic nulling from older studies. Some measurements may be subject to misclassifying amplitude modulation for nulling due to the limited sensitivity of the observations. Additionally, as there is no standard for classifying quasiperiodic nulling, multiple authors may have their own definitions. Due to these caveats, we exercise caution before drawing strong conclusions and urge a larger study to test quasiperiodicities in a larger sample of nulling pulsars using a standard. Such a cohesive study can lead to conclusions free of biases.

\subsection{Nulling Quasiperiodicity and Other Pulsar Parameters}

We compared the nulling quasi-periodicity with other parameters to test possible correlations. We find no correlations between the nulling quasi-periodicity and spin period; however, we observe a weak correlation between the nulling periodicity and nulling fraction. Given the fragility of the correlation, we choose not to draw any conclusions until a stronger relationship is observed.

\subsection{Origins of Nulling}

Nulling has been observed to be a broadband feature; however, the fraction of nulling appears to be able to change significantly between observations, e.g. as is the case for J1901$-$0906 and J1136$+$1551. Generally, we find the nulling fraction measurements for a majority of the pulsars to be consistent throughout frequency between various observations. This has also been observed in the past via multifrequency observations of nulling pulsar \citep[e.g.][]{2007Bhat, 2017Naidu}. All frequencies being affected equally implies the phenomenon responsible for nulling does not affect the geometry of the beam. Instead, a consistent nulling fraction through frequencies implies a magnetospheric shift that completely dims, ceases or redirects the emission.

For arguments suggesting the nulling phenomenon to be a weaker, non-observable emission mode rather than a cessation of emission, multifrequency measurements of nulling are equivocal. Although pulse emission is brighter at lower frequencies (negative spectral index), instruments at lower frequencies are also less sensitive. These effects negate each other, limiting conclusions from multifrequency nulling studies.
%merge 

%Something yet to be explored in this space is 
We have not yet explored the polarisation behaviour of pulsars before and after nulls. Polarisation has rarely been connected to nulling, and has been suggested by \citet{Sheikh2021}. Since there is no observable emission during a null, these parameters do not link in the conventional sense. However, studying the polarisation of pulses before and after a null, and comparing it to other pulses, may help us learn about the origins of nulls. If nulling is a magnetospheric change, as is supported by our findings, and the change occurs on a timescale longer than two periods, we should be able to observe changes within the polarisation. Alternatively, if there is no change in polarisation adjacent to nulls, we can constrain the timescale of the phenomenon that causes nulling, which can constrain the physics of the phenomenon.

\section{Conclusion}

Even after years of investigations, the nulling phenomenon in pulsar radio emission remains poorly understood, and the degree of nulling in many pulsars is often inaccurately quantified. Although nulling is generally measurable and characterised in detail in brighter pulsars, for the vast majority of pulsars, measuring and categorising nulling becomes difficult due to low S/N observations. Additionally, even in pulsars where nulling is measurable, in most cases, mainly the nulling fraction is reported, whereas the nulling duration and (possible) quasi-periodicity are also needed to fully understand the nulling behaviour. To address these issues, we have developed and explored the applicability of a new algorithm that uses sums of pulse fluence to better measure nulling in low S/N pulsars. We also measured and analysed the quasi-periodicity for pulsars in our sample.

The probabilities of picking a bursting pulse and a nulling pulse at random are $(1-n_{\rm F})$ and $n_{\rm F}$, respectively. If we sum $\mathbb N$ pulses, then by using combinatorics, we can find the probability of picking $z$ nulling pulses out of $\mathbb N$ total pulses.
%can be described by Equation \ref{eq:comb}. 
In doing this, we also assume the intrinsic pulse fluence and noise are normally distributed; hence, the summation of their samples will also be normally distributed. Since we know the probability of the outcomes from summing $\mathbb N$ pulses, which is dependent on $n_{\rm F}$, and we know the distribution of the sums from each outcome, using the distribution of the sums from $\mathbb N$ random pulses, we can measure $n_{\rm F}$. 

Using this approach, we tested our method for $\mathbb N=1 \,\,\&\,\, \mathbb N=3$ and compared it to the GMM \citep{Kaplan2018, Anumarlapudi2023}. We find $\mathbb N=1$ and GMM methods produced similar results as they are similar methodologies. Although their performances differ for weak pulses in simulated data, they show similar results for real pulsars. The only places where they differed for real pulsars data were in cases where the pulse fluences are not distributed by a single Gaussian. $\mathbb N=3$ measurements generally agree with the other methods, differing by a large margin for weak pulsars and again for cases where pulse fluences are not distributed by a single Gaussian. For weaker pulsars that have been reported to have low nulling fractions, the $\mathbb N=3$ method measures values closer to the literature values.

Using a fairly robust methodology, we also attempted to measure quasi-periodicity in nulling for all pulsars in our sample. We find five pulsars show clear quasiperiodicity, three of which have not been previously reported to show this behaviour, while the other pulsars in our sample do not show any signs of nulling quasi-periodicity. We also gather the largest sample of quasiperiodically nulling pulsars to comment on past work. We find that quasiperiodically nulling pulsars may not be bound to the death valley, contrary to the findings of \cite{Grover2024a}, and $\dot E$ may be used to distinguish between different types of periodicities.

\begin{acknowledgments}

%MRO/Pawsey/OzSTAR
This scientific work uses data obtained from \textit{Inyarrimanha Ilgari Bundara}, the CSIRO Murchison Radio-astronomy Observatory. We acknowledge the Wajarri Yamaji People as the Traditional Owners and native title holders of the Observatory site. Establishment of CSIRO's Murchison Radio-astronomy Observatory is an initiative of the Australian Government, with support from the Government of Western Australia and the Science and Industry Endowment Fund. Support for the operation of the MWA is provided by the Australian Government (NCRIS), under a contract to Curtin University administered by Astronomy Australia Limited. This work was supported by resources provided by the Pawsey Supercomputing Research Centre with funding from the Australian Government and the Government of Western Australia.
%, as well as by resources awarded under Astronomy Australia Ltd's ASTAC merit allocation scheme on the OzSTAR national facility at the Swinburne University of Technology. The OzSTAR program receives funding in part from the Astronomy National Collaborative Research Infrastructure Strategy (NCRIS) allocation provided by the Australian Government.
%RTP
G.G. is supported by an Australian Government Research Training Program (RTP) Stipend and RTP Fee-Offset Scholarship.
%Facilities
%GMRT

\end{acknowledgments}
\appendix
% \begin{figure}[htb!]
%     \centering
%     \includegraphics[width=0.7\linewidth]{Simulated_nsum3_0.1_50_histogram_fit_3sum.png}
%     \includegraphics[width=0.7\linewidth]{Simulated_nsum3_0.1_50_rerun_histogram_fit_3sum.png}
%     \caption{Similar figures to the right panel of Figures \ref{fig: J1919+0021_example} and \ref{fig: J1456-6843_example}. These Gaussian mixture is generated by the parameters estimated by $\mathbb N=3$ for the simulated fluences of $\mu_{\rm P}=0.1,n_{\rm F}=50\%$. The top figure is generated with the estimated $n_{\rm F}=30\%$}
%     \label{fig:histfit}
% \end{figure}
% \section{Supplementary Figures}
\begin{figure}[htpb!]
    \centering
    \includegraphics[width=\linewidth]{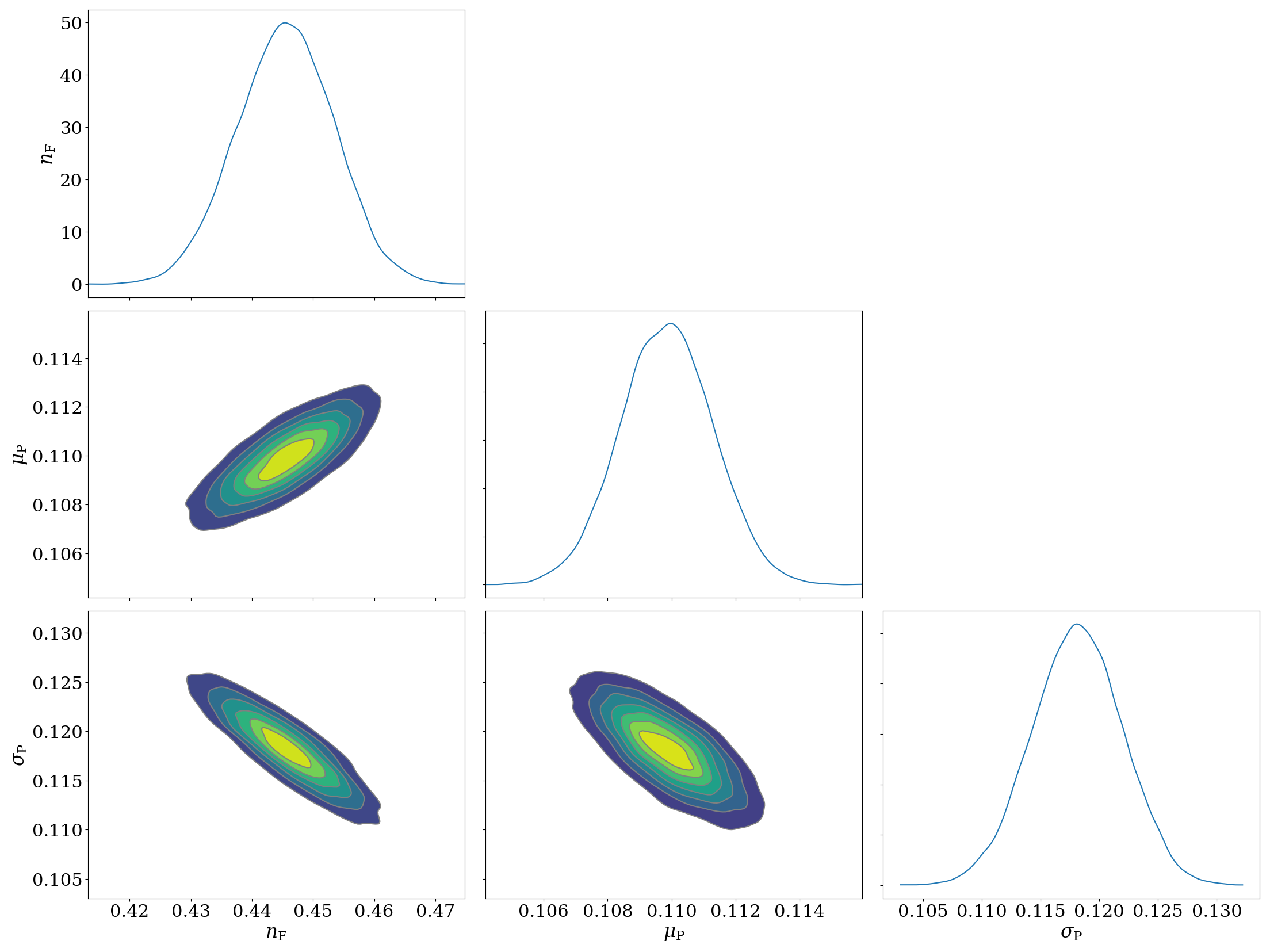}
    
    \caption{This is a corner plot produced by $\mathbb N=3$ for the simulated parameters $\mu_{\rm P}=0.1, n_{\rm F}=50\%$.}
    \label{fig:nsumcorner}
\end{figure}
\begin{figure}[htp!]
    \centering
    \includegraphics[width=0.8\linewidth]{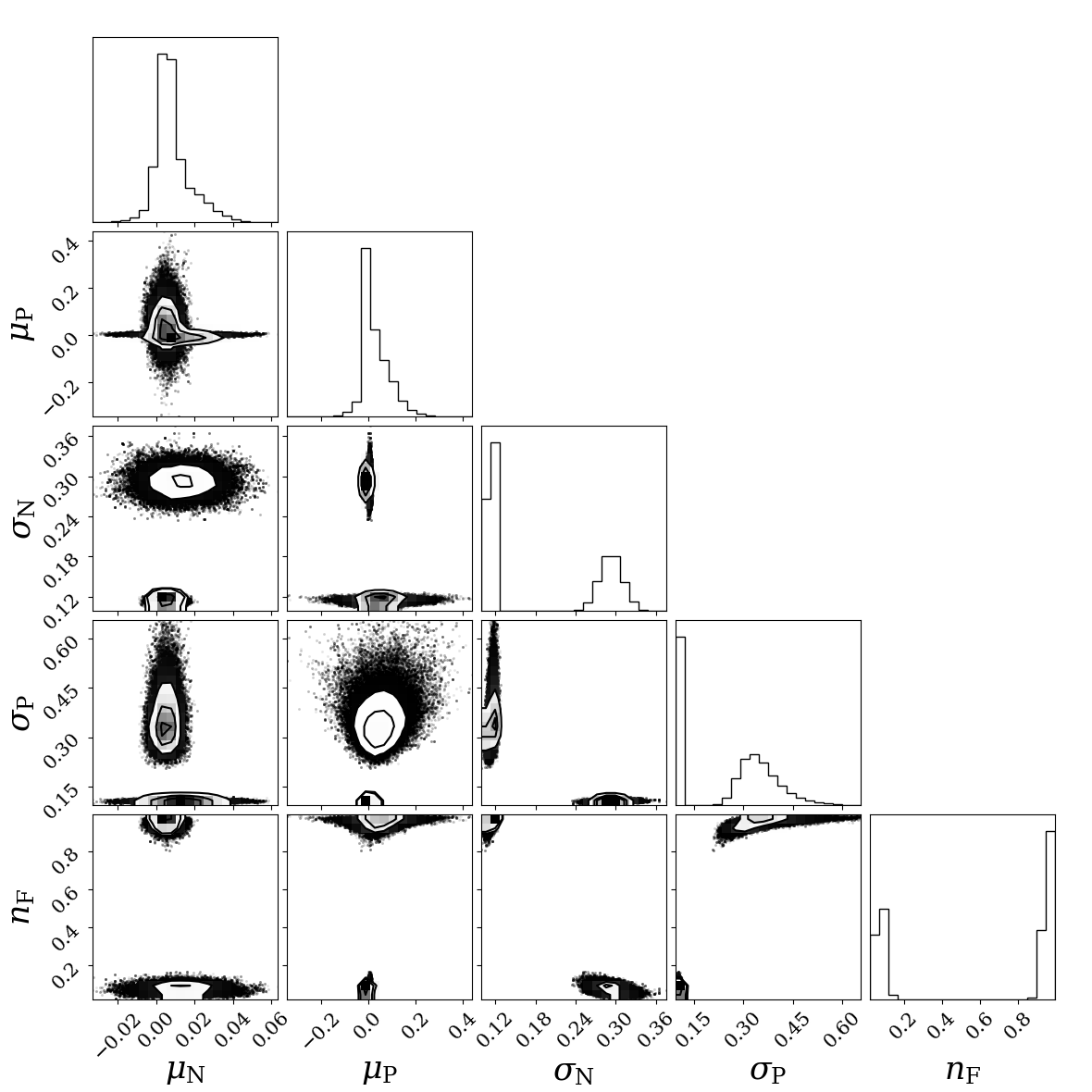}
    \caption{A corner plot produced by GMM for the case of $\mu_{\rm P}=0.001, n_{\rm F}=10\%$. The bimodal nature of the posteriors is very clear in the figure for $\sigma$ and $n_{\rm F}$.}
    \label{fig:cornerplot example}
\end{figure}
\bibliography{sample631}
\end{document}